\documentclass[a4paper]{article}

\usepackage{geometry}
\usepackage{amsfonts}
\usepackage{cite}
\usepackage{amssymb}
\usepackage{bm}
\usepackage{graphicx}
\usepackage{natbib}
\usepackage{titlesec}
\titleformat{\section}{\large\bfseries}{\thesection}{1em}{}
\title{A partial least squares algorithm handling ordinal variables \\also in presence of a small number of categories}

\author{Gabriele Cantaluppi\\Dipartimento di Scienze statistiche\\Universit\`a Cattolica del Sacro Cuore\\Largo A. Gemelli, 1, 20123 Milano, Italy\\gabriele.cantaluppi@unicatt.it}

\date{}

\begin{document}
\thispagestyle{empty}

\includegraphics[scale=.99]{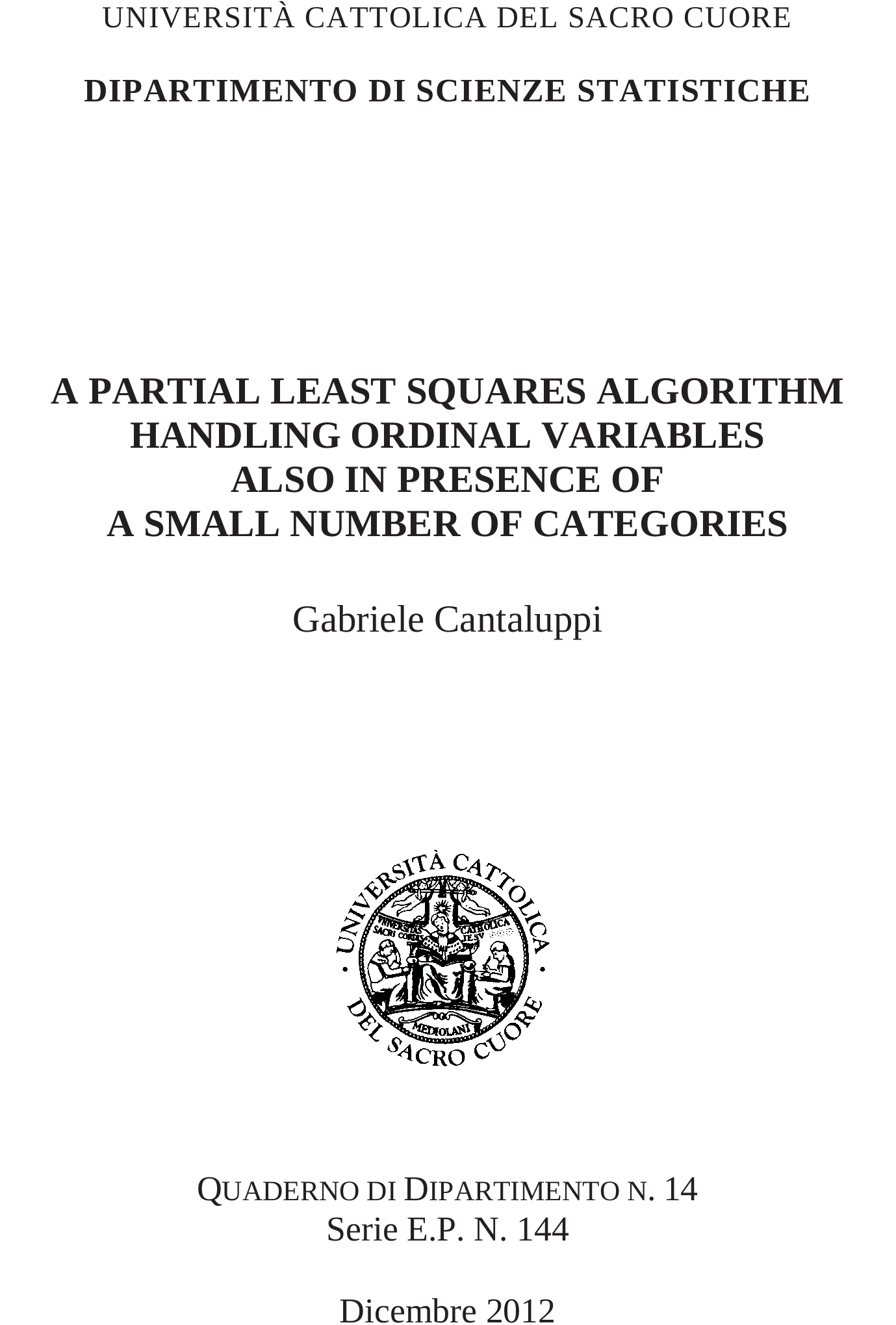}

\newpage
~
\thispagestyle{empty}

\newpage
{\bf\begin{center}
{\Large A Partial Least Squares Algorithm}
\end{center}
\begin{center}
{\Large Handling Ordinal Variables also in Presence}
\end{center}
\begin{center}
{\Large of a Small Number of Categories}
\end{center}
}

\vspace{9pt}
\begin{center}
{\Large Gabriele Cantaluppi}

\vspace{9pt}
Dipartimento di Scienze statistiche\\Universit\`a Cattolica del Sacro Cuore\\Largo A. Gemelli, 1, 20123 Milano, Italy\\gabriele.cantaluppi@unicatt.it
\end{center}

\vspace{18pt}
\noindent{\bf Abstract: }
The partial least squares (PLS) is a popular modeling technique commonly used in social sciences.
The traditional PLS algorithm deals with variables measured on interval scales while data are often collected on ordinal scales:
a reformulation of the algorithm, named ordinal PLS (OPLS), is introduced, which properly deals with ordinal variables.
An application to customer satisfaction data and some simulations are also presented.
The technique seems to perform better than the traditional PLS when the number of categories of the items in the questionnaire is small (4 or 5) which is typical in the most common practical situations.

\vspace{12pt}
\noindent{\bf Keywords:} Ordinal Variables, Partial Least Squares, Path Analysis, Polychoric Correlation Matrix, Structural Equation Models with Latent Variables

\section{Introduction}

The partial least squares (PLS) technique is largely used in socio-economic studies where path analysis is performed with reference to the so-called structural equation models with latent variables (SEM-LV).

Furthermore, it often happens that data are measured on ordinal scales; a typical example concerns customer satisfaction surveys, where responses given to a questionnaire are on Likert type scales assuming a unique common finite set of possible categories.

In several research and applied works, averages, linear transformations, covariances and Pearson correlation coefficients are conventionally computed also on the ordinal variables coming from surveys. This practice can be theoretically justified by invoking the \textit{pragmatic} approach to statistical measurement \citep{Hand_2009}. Namely according to this approach 'the precise property being measured is defined simultaneously with the procedure for measuring it' \citep{Hand_2012}, so when defining a construct, e.g. the overall customer satisfaction, the measuring instrument is also defined, and 'in a sense this makes the scale type the choice of the researcher' \citep{Hand_2009}.

A more traditional approach \citep[see][]{Stevens_1946} would require appropriate procedures to be adopted in order to handle manifest indicators of the ordinal type. Within the well-known covariance-based framework, several approaches are suggested in order to appropriately estimate a SEM-LV model; in particular, \cite{muthen84}, \cite{jore-ordinal} and \cite{Bollen_1989} make the assumption that to each manifest indicator there corresponds an underlying continuous latent variable, see Section \ref{sectSomeassumptionsonthegenesisofordinalcategoricalobservedvariables}.

Other approaches have been proposed to deal with ordinal variables within the Partial Least Squares (PLS) framework for SEM-LV: \cite{Jakobowicz2007} base their procedure on the use of generalized linear models; \cite{Nappo} and \cite{LauroNappoGrassiaMiele} on Optimal Scaling and Alternating Least Squares; \cite{RussolilloLauro} on the Hayashi first quantification method \citep{Hayashi}.

As observed by \cite{RussolilloLauro} in the procedure by \cite{Jakobowicz2007} a value is assigned to the impact of each explanatory variable on each category of the response, while the researcher may be interested in the impact of each explanatory variable on the response as a whole. The same issue characterizes the techniques illustrated in the Chapter 5 by \cite{Lohmoeller_1989}.
The present proposal goes in this direction: a reformulation of the PLS path modeling algorithm is introduced, see Section \ref{AnequivalentreformulationofthePLSalgorithmanditsimplementationwithordinalvariables}, allowing us to deal with variables of the ordinal type in a manner analogous to the covariance based procedures.

In this way we recall the traditional psychometric approach, by applying a method for treating ordinal measures according to the well-known \cite{Thurstone_1959} scaling procedure, that is assuming the presence of a continuous underlying variable for each ordinal indicator.
The polychoric correlation matrix can be defined. We show that by using this matrix one can obtain parameter estimates also within the PLS framework.

When the number of points of the scale is sufficiently high the value of the polychoric correlation between two variables is usually quite close to that of the Pearson correlation; in these situations there would be no need to have recourse to polychoric correlations and the traditional PLS algorithm may be applied. However, to make the response of interviewed people easier, it is common practice to administer questionnaires whose items are measured on at most 4 or 5 point alternatives: in these circumstances the proposed modification of the PLS algorithm seems to be appropriate.

An application to customer satisfaction data and some simulations conclude the paper.

\section{The Structural Equation Model with latent variables}

A linear SEM-LV consists of two sets of equations \citep[see][]{Bollen_1989}: the structural or inner model describing the path of the relationships among the latent variables, and the measurement or outer model, representing the relationships linking the latent variables, which cannot be directly observed, to appropriate corresponding manifest variables.

\vspace{.25cm}
\noindent\textbf{The inner model}

\vspace{.125cm}
\noindent The structural model is represented by the following linear relation
\begin{equation}
\label{equation010}
\bm\eta =\mathbf B \bm\eta+\bm\Gamma\bm\xi+\bm\zeta
\end{equation}
where $\bm{\eta}$ is an $(m \times 1)$ vector of latent endogenous random variables (dependent variables); $\bm{\xi}$ is an $(n \times 1)$ vector of latent exogenous random variables; $\bm{\zeta}$ is an $(m \times 1)$ vector of error components, zero mean random variables. $\mathbf{B}$ and $\bm{\Gamma}$ are respectively $(m \times m)$ and $(m \times n)$ matrices containing the so-called structural parameters. In particular the matrix $\mathbf{B}$ contains information concerning the linear relationships among the latent endogenous variables: their elements represent the direct casual effect on each $\eta_i$ $(i=1,\dots,m)$ of the remaining $\eta_j$ $(j\ne i)$. The matrix $\bm{\Gamma}$ contains the coefficients explaining the relationships among the latent exogenous and the endogenous variables: their elements represent the direct causal effects of the $\bm{\xi}$ components on the $\eta_i$ variables.
\\When the matrix $\mathbf{B}$ is lower triangular or can be recast as lower triangular by changing the order of the elements in $\bm{\eta}$ (which is possible if $\mathbf{B}$ has all zero eigenvalues \cite[see e.g.][]{Faliva}) and $\bm{\Psi}$ is diagonal, then the model (\ref{equation010}) is said to be causal or of the recursive type, which excludes feedback effects. In the sequel we will assume $\mathbf{B}$ to be lower triangular.

\vspace{.25cm}
\noindent\textbf{The outer model}

\vspace{.125cm}
\noindent We consider only measurement models of the reflective type, which are adopted \citep[see][]{Diamantopoulos2008} when the latent variables 'determine' their manifest indicators, and are defined according to the following linear relationships
\begin{eqnarray}
\label{equation040}
\mathbf x=\bm\Lambda_X\bm\xi+\bm\varepsilon_X \\
\label{equation045}
\mathbf y=\bm\Lambda_Y\bm\eta+\bm\varepsilon_Y
\end{eqnarray}
where the vector random variables $\mathbf{x}$ $(q \times 1)$ and $\mathbf{y}$ $(p \times 1)$ represent the indicators for the latent variables $\bm{\xi}$ and $\bm{\eta}$, respectively.
Each latent construct $\eta_i$ in $\bm\eta$ and $\xi_i$ in $\bm\xi$ is characterized by a set of indicators, $Y_{ih}, h=1,\dots,p_i$ for $\eta_i$ and $X_{ih}, h=1,\dots,q_i$ for $\xi_i$.

The assumption of uncorrelation among the error components and uncorrelation between the errors and the independent variables in relationships (\ref{equation010})-(\ref{equation045}) is made. All latent variables and errors are usually assumed to be multivariate distributed according to a Normal random variable.

Measurement models of the formative and of the MIMIC type may also be defined \citep[see][]{EspositoVinzi_2010} but are not considered here.

It often happens that the indicators $Y_{ih}, X_{ih}$ are measured on ordinal scales, e.g. the responses given by the respondents to a questionnaire are on Likert type scales that assume a unique common finite set of possible categories. In this instance appropriate procedures are adopted for parameter estimation in SEM-LV in order to treat manifest indicators of the ordinal type. In \cite{muthen84}, \cite{jore-ordinal} \cite{Bollen_1989} and \cite{BollenOlivares2007} estimation procedures within a covariance-based framework are presented, which are based on the assumption that for each manifest indicator there corresponds a further underlying continuous latent variable, whose definition is here described in Section \ref{sectSomeassumptionsonthegenesisofordinalcategoricalobservedvariables}.

\vspace{.25cm}
\noindent\textbf{The PLS specification}

\vspace{.125cm}
\noindent Observe that the structural relationship (\ref{equation010}) can be re-written in matrix notation as
\begin{equation}
\label{equation020}
\left[ 
\begin{array}{c}
\bm{\xi} \\ 
\bm{\eta}
\end{array}
\right]
=
\left[ 
\begin{array}{cc}
\mathbf{I} & \mathbf O \\ 
\bm{\Gamma} & \mathbf B
\end{array}
\right]
\left[ 
\begin{array}{c}
\bm{\xi} \\ 
\bm{\eta}
\end{array}
\right]
+
\left[ 
\begin{array}{c}
\mathbf 0\\ 
\bm{\zeta}
\end{array}
\right]
\end{equation}
and the reflective measurement model (\ref{equation040})-(\ref{equation045}) as
$$
\left[ 
\begin{array}{c}
\mathbf x \\ 
\mathbf y
\end{array}
\right]
=
\left[ 
\begin{array}{cc}
\bm\Lambda_X & \mathbf 0 \\ 
\mathbf 0    & \bm\Lambda_Y
\end{array}
\right]
\left[ 
\begin{array}{c}
\bm\xi \\ 
\bm\eta
\end{array}
\right]
+
\left[ 
\begin{array}{c}
\bm\varepsilon_X \\ 
\bm\varepsilon_Y
\end{array}
\right].
$$
In the initial works on PLS \citep{Wold_1985, Lohmoeller_1989} the same notation is used for endogenous and exogenous entities; thus the above relationships are re-written as
\begin{eqnarray}
\label{equation030}
\bm{Y}=\mathbf D \bm{Y}+\bm{\nu} \\
\label{equation050}
\bm{X}=\bm{\Lambda Y}+\bm\varepsilon
\end{eqnarray}
by having (re-)defined
$$\bm Y\equiv
\left[
\begin{array}{c}
\bm\xi \\ 
\bm\eta
\end{array}
\right],
\ 
\mathbf{D}\equiv
\left[
\begin{array}{cc}
\mathbf{I} & \mathbf O \\ 
\bm{\Gamma} & \mathbf B
\end{array}
\right],
\ 
\bm\nu\equiv
\left[
\begin{array}{c}
\mathbf 0 \\ 
\bm\zeta
\end{array}
\right],
\ 
\bm X\equiv\left[ 
\begin{array}{c}
\mathbf x \\ 
\mathbf y
\end{array}
\right],
\ 
\bm\Lambda\equiv
\left[ 
\begin{array}{cc}
\bm\Lambda_X & \mathbf 0 \\ 
\mathbf 0    & \bm\Lambda_Y
\end{array}
\right],
\ 
\bm\varepsilon\equiv\left[ 
\begin{array}{c}
\bm\varepsilon_X \\ 
\bm\varepsilon_Y
\end{array}
\right].
$$
The sub-matrix $\mathbf B$, corresponding to the matrix $\mathbf B$ which appears in (\ref{equation010}), is assumed to be lower triangular. The so-called Wold predictor specification, $E(\zeta_i|\eta_1,\dots,\eta_{i-1})=0, i=1,\dots,m$, is made, giving rise to structural equation models of the causal type.
\\The measurement model of the reflective type is named 'Mode A' in the PLS terminology.

\section{Assumptions on the genesis of ordinal categorical observed variables}
\label{sectSomeassumptionsonthegenesisofordinalcategoricalobservedvariables}

The set of responses are assumed to be expressed on a conventional ordinal scale. This type of scale requires, according to the classical psychometric approach, appropriate methods to be applied.
Here we propose to adopt a traditional psychometric approach, by considering the existence of an underlying continuous unobservable latent variable for each observed ordinal manifest variable.

Following the PLS notation (\ref{equation030}) and (\ref{equation050}) for structural equation models with latent variables, the set of responses gives rise to a $K$-dimensional random categorical variable, say $\bm X=(X_1, \dots, X_K)'$, whose components may assume, for simplicity, the same $I$ ordered categories, denoted by the conventional integer values $i=1, \dots, I$.
$K$ is the dimension of $\bm X$ in (\ref{equation050}), corresponding to $q+p$ in (\ref{equation040})-(\ref{equation045}), the sum of the dimensions of $\mathbf x$ and $\mathbf y$.

Let $P(X_k=i)=p_{ki}$, with $\sum^I_{i=1} p_{ki}=1, \forall k$, be the corresponding marginal probabilities and let
\begin{equation}
\label{Zan_01}
F_k(i)=\sum_{j\le i} p_{kj}
\end{equation}
be the probability of observing a conventional value $x_k$ for $X_k$ not greater than $i$.
Furthermore assume that to each categorical variable $X_k$ there corresponds an unobservable latent variable $X^*_k$, which is represented on an interval scale with a continuous distribution function $\Phi_k(x^*_k)$.
The distribution for the continuous $K$-dimensional latent random variable $\bm X^*=(X^*_1,\dots,X^*_K)$ is usually assumed to be multinormal.
Each observed ordinal indicator $X_k, k=1,\dots,K,$ is related to the corresponding latent continuous $X^*_k$ by means of a non linear monotone function \citep[see][]{Bollen_1989,muthen84,jore-ordinal}, of the type
\begin{equation}
\label{ZBBCequation003}
X_k=\left\{
\begin{array}{l}
1 \quad \quad \ \mathrm{if} \quad X^*_k \le a_{k,1} \\
2 \quad \quad \ \mathrm{if} \quad a_{k,1} < X^*_k \le a_{k,2} \\
\vdots \\
I_k-1 \ \ \mathrm{if} \quad a_{k,I_k-2} < X^*_k \le a_{k,I_k-1} \\
I_k \quad \quad \ \mathrm{if} \quad a_{k,I_k-1} < X^*_k \\
\end{array}
\right.
\end{equation}
where $a_{k,1},\dots,a_{k,I_k-1}$ are marginal threshold values defined as $a_{k,i}=\Phi^{-1}(F_k(i)), i=1,\dots,I_k-1$, being $\Phi(\cdot)$ the cumulative distribution function of a specific random variable, usually the standard Normal, \cite{jore-ordinal}; $I_k\le I$ denotes the number of categories effectively used by the respondents; $I_k=I$ when each category has been chosen by at least one respondent.

We also set $a_{k,0}=-4$ and $a_{k,I_k}=4$ and set to $-4$ or 4 threshold values respectively lower than $-4$ or larger than 4.

\section{Appropriate covariance matrix in presence of ordinal categorical variables}
\label{sectDerivingaappropriatecovariancematrixinpresenceofordinalcategoricalvariables}

We remember that covariance-based estimation procedures look for the parameter values minimizing the distance between the covariance matrix of the manifest variables, specified as a function of the parameters, and its sample counterpart. Since, in case of ordinal variables it is not possible to compute the covariance matrix, we have recourse to the polychoric correlation matrix or the polychoric covariance matrix \citep[see][]{Bollen_1989}.

For two generic ordinal categorical variables $X_h$ and $X_k$, $h,k \in \{1,\dots,K\}$, the \textit{polychoric} correlation is defined \citep[see][]{Drasgow_1986,Olsson_1979}, as the value of $\rho$ maximizing the loglikelihood typically conditional on the marginal threshold estimates
$$
\sum_{i=1}^{I_h}\sum_{j=1}^{I_k} n_{ij}\ln(\pi_{ij}),
$$
where $n_{ij}$ is the number of observations for the categories $i$th of $X_h$ and $j$th of $X_k$,
$$
\pi_{ij}=\Phi_2(a_{h,i},a_{k,j})-\Phi_2(a_{h,i-1},a_{k,j})-\Phi_2(a_{h,i},a_{k,j-1})+\Phi_2(a_{h,i-1},a_{k,j-1}),
$$
being $\Phi_2(\cdot)$ the standard bivariate Normal distribution function with correlation $\rho$ conditional on the threshold values, $a_{h,i},a_{k,j}$, for $X_h$ and $X_k$, respectively estimated by having recourse e.g. to the two marginal latent standard Normal variates according to the usual two step computation, $a_{k,0}=-\infty$ and $a_{k,I_k}=+\infty$.
\\Later on we will assume that $I_k=I$, that is each category has been chosen by at least one respondent, possibly substituting a negligible quantity, e.g. $\varepsilon=0.5$, to the zero $n_{ij}$s.

By considering the polychoric coefficients for each pairs $X_h$ and $X_k$, $h,k \in \{1,\dots,K\}$ we can obtain the polychoric correlation matrix (and also the covariance matrix, if appropriate location and scale values are assigned to the underlying latent variables $X^*_k$ \citep{jore-ordinal}), which, according to the covariance-based approach, is necessary for the parameter estimation of a structural model with manifest indicators of the ordinal type.

In case of manifest variables of generic type appropriate correlations should be computed \citep[see][]{Drasgow_1986}; in particular: a) polychoric correlation coefficients for pairs of ordinal variables, b) polyserial correlation coefficients between an ordinal variable and one defined on an interval or ratio scale, and c) Pearson correlation coefficients between variables defined on interval or ratio scales. However we will assume, later on, that only variables on ordinal scales are present.

\section{Application within the PLS framework}

In presence of manifest indicators of the ordinal type, we suggest a slightly modified version of model (\ref{equation030})-(\ref{equation050}), where the manifest variables $\bm X$ in relationship (\ref{equation050}) are in a certain sense 'replaced' by the underlying unobservable latent variables $\bm X^*$.
\\We do not write explicitly the dependence between $\bm X$ and $\bm X^*$, since for the subject $s=1,2,\dots,N$ the real point value $x_{ks}^*$ for each indicator $X_k^*$ is not known: we only assume that it belongs to the interval defined by the threshold values in (\ref{ZBBCequation003}) having as image the observed category $x_{ks}$.
\\It will be possible to obtain point estimates for the parameters in $\mathbf D$ and $\bm\Lambda$, while only estimates of the threshold values will be directly predicted with regard to the scores of the latent variables $Y_j, j=1,\dots,n+m$.
\\The PLS algorithm structure typically adopted to obtain the PLS estimates in presence of standardized latent variables is briefly presented. Mode A (reflective outer model) is considered for outer estimation of latent variables and the centroid scheme for the inner estimation \citep{Tenenhaus2005, Lohmoeller_1989}.\\Some linear algebra restatement of the algorithm is necessary, which renders the usual procedure apt to be applied with minor changes also in presence of manifest variables of the ordinal type.

\subsection{The PLS algorithm}

The PLS procedure obtaining the estimates of the parameters in (\ref{equation030})-(\ref{equation050}) consists of a first iterative phase which produces the latent score estimates; subsequently the values of the vector $\bm{\theta}$, containing all the unknown parameters in the model ($\mathbf{D}$, $\bm{\Gamma}$ etc.), are estimated, by applying the Ordinary Least Squares method to all the linear multiple regression sub-problems into which the inner and outer models are decomposed.

Remember that the whole set of true latent variables (always measured as differences from their respective average values) is summarized by the vector
$$
\bm{Y}= \left[ Y_1,\dots,Y_n,Y_{n+1},\dots,Y_{n+m}\right]'
$$
being the first $n$ elements of the exogenous type and the remaining $m$ endogenous.
Observe that, since we are in presence of models of the causal type, the generic endogenous variable $Y_j$, $j=n+1,\dots,n+m$, may only depend on the exogenous variables $Y_1,\dots,Y_n$ and on a subset of its preceding endogenous latent variables.

With reference to the inner model, a square matrix $\mathbf{T}$, of order $(n+m)$, indicating the structural relationships among latent variables may be defined:
\begin{equation}
\label{equationsquarematrixT}
\begin{tabular}{cc|cccccc|}
\cline{3-8}
\rule[0.1cm]{0cm}{.4cm}& $Y_{1}$ & 0 & 0 & 0 & 0 & 0 & 0 \\ 
$\mathrm{exogenous}$ & \small{\vdots} & 0 & 0 & 0 & 0 & 0 & 0 \\ 
\rule[-.2cm]{0cm}{.65cm}& $Y_{n}$ & 0 & 0 & 0 & 0 & 0 & 0 \\ \cline{3-5}
\rule[-.2cm]{0cm}{.65cm}& $Y_{n+1}$ &  & \multicolumn{1}{|c}{} & \multicolumn{1}{|c}{} & \multicolumn{1}{|c}{0} & 0 & 0 \\ \cline{3-6}
$\mathrm{endogenous}$ & \small{\vdots} &  & \multicolumn{1}{|c}{} & \multicolumn{1}{|c}{} & \multicolumn{1}{|c}{} & \multicolumn{1}{|c}{0} & 0 \\ \cline{3-7}
\rule[-.2cm]{0cm}{.65cm}& $Y_{n+m}$ &  & \multicolumn{1}{|c}{} & \multicolumn{1}{|c}{} & \multicolumn{1}{|c}{} & \multicolumn{1}{|c}{} & \multicolumn{1}{|c|}{0} \\ 
\cline{3-7}\cline{3-8}
\end{tabular}
=\mathbf T
\end{equation}
The generic element $t_{jk}$ of $\mathbf{T}$ is given unit value if the endogenous $Y_j$ is directly linked to $Y_k$; $t_{jk}$ is null otherwise. Then $\mathbf T$ may be defined as the indicator matrix of $\mathbf D$ in (\ref{equation030}) by having set to 0 the elements on the main diagonal.

The PLS algorithm  follows an iterative procedure, which defines, at the $r$th generic step, the scores of each latent variable $Y_j$, according the so-called 'outer approximation', as a standardized linear combination of the manifest variables corresponding to $Y_j$
\begin{equation}
\label{boaricantaluppi.equation100}
\hat Y_j=\sum_{h=1}^{p_j} w_{jh}^{(r)}(X_{jh} - \bar x_{jh}), \quad j=1,2,\dots,n+m
\end{equation}
with appropriate weights $w_{j1}^{(r)},\dots, w_{jp_j}^{(r)}$ summing up to 1 \citep[see][]{Lohmoeller_1989}.
\\In the PLS framework each latent variable is thus defined as a 'composite' of its manifest indicators.

\vspace{.125cm}
\noindent\textit{Step 0.} The starting step of the algorithm uses an arbitrarily defined weighting system that, for the sake of simplicity, may be set to
\begin{equation}
\label{boaricantaluppi.equationpesiinizialimod}
\{w_{jh}^{(0)}=1/p_j, \quad h=1,2,\dots,p_j\} \quad j=1,2,\dots,n+m.
\end{equation}
The initial scores of the latent variables $Y_j$, $j=1,2,\dots,n+m$, are defined as linear combinations of the centered values of the corresponding manifest variables $X_{jh}$, $h = 1,2,\dots,p_j$; thus for the generic subject $s$, $s=1,2,\dots,N$, we have:
\begin{equation}
\label{equation110}
\hat y_{js}= \sum^{p_j}_{h=1} w_{jh}^{(0)} (x_{jhs} - \bar x_{jh})
\end{equation}
where $\bar x_{jh}$, $h=1,2,\dots,p_j$, denote the mean of the manifest variables $X_{jh}$ associated to the latent variable $Y_j$. Observe that at step 0 the weights sum up to 1 by definition. \\The scores are then standardized
\begin{equation}
\label{latstand}
\tilde y_{js}=\hat y_{js}f_j
\end{equation}
where $$f_j=\left(\frac{1}{N-1}\sum_{s=1}^Ny_{js}^2\right)^{-\frac{1}{2}}$$ is an estimate of the variance of $\hat Y_j$, being $N$ the number of available observations.
\\We then set
\begin{equation}
\label{latstand1}
\hat y_{js}=\tilde y_{js}.
\end{equation}

\vspace{.25cm}
\noindent\textbf{Iterative phases of the PLS algorithm}

\vspace{.125cm}
\noindent\textit{Step 1.} Define for each latent variable $Y_j$ an instrumental variable $Z_j$ as a linear combination of the estimates of the latent variables $Y_k$ directly linked to $Y_j$
\begin{equation}
\label{equation130}
Z_j =  \sum^{n+m}_{k=1}\tau_{jk}\hat Y_k,
\end{equation}
where
\begin{equation}
\label{equation133}
\tau_{jk} = \max(t_{jk},t_{kj}) \mathrm{sign}[Cov(\hat Y_j, \hat Y_k)]
\end{equation}
(remember that $t_{jk}$ is the generic element of the matrix $\mathbf T$, used to specify the relationships in the inner model: $t_{jk} = 1$ if the latent variable $Y_j$ is connected with $Y_k$ in the path model representation, $t_{jk} = 0$ otherwise, see (\ref{equationsquarematrixT})) and
\begin{equation}
\label{equation135}
Cov(\hat Y_j, \hat Y_k) = \frac{1}{N - 1} \sum^N_{s=1} \hat y_{js} \hat y_{ks}
\end{equation}
having $\hat Y_j$ zero mean.

\vspace{.25cm}
\noindent \textit{Step 2.} In case of the so-called Mode A (reflective outer model), at every stage $r$ of the iteration ($r=1,2,\dots$) update the vectors of the weights $w_j^{(r)}$ as
\begin{equation}
\label{boaricantaluppi.equation16013}
w_{jh}^{(r)} = \pm C_{jh} /  \sum^{p_j}_{h=1}C_{jh} ,  \quad j = 1,2,\dots,n+m, \ h = 1,2,\dots,p_j,
\end{equation}
where
\begin{eqnarray}
\label{boaricantaluppi.equation16014}
C_{jh} = Cov(X_{jh}, Z_j) = \frac{1}{N - 1} \sum^N_{s=1} (x_{jhs} - \bar x_{jh})(z_{js} - \bar z_{jh}),
\\
\label{boaricantaluppi.equation16015}
\bar z_{jh} = \frac{1}{N} \sum^N_{s=1} z_{js},   \quad  \pm = sign( \sum^{p_j}_{h=1}sign[Cov(X_{jh},\hat Y_j)]),
\end{eqnarray}
being
$$
Cov(X_{jh}, \hat Y_j) = \frac{1}{N - 1} \sum^N_{s=1} (x_{jhs} - \bar x_{jh})(\hat y_{js} - \bar y_{j})
\quad \quad \mathrm{and} \quad \quad
\bar y_{j} = \frac{1}{N} \sum^N_{s=1} \hat y_{js}.
$$

\vspace{.25cm}
\noindent \textit{Step 3.} Update the outer approximation:
\begin{equation}
\label{boaricantaluppi.equation16018}
\hat y_{js} = \frac{1}{\sum^{p_j}_{h=1}w_{jh}^{(r)}} \sum^{p_j}_{h=1} w_{jh}^{(r)} (x_{jhs} - \bar x_{jh}),
\end{equation}
and standardize as in (\ref{latstand})-(\ref{latstand1}).

\vspace{.25cm}
\noindent \textit{Looping.} Loop step 1 to step 3 until the following convergence stop criterion is attained
$$
\left\{ \sum^{m+n}_{j=1} \sum^{p_j}_{h=1} \left( w_{jh}^{(r)} - w_{jh}^{(r-1)} \right)^2\right\}^{1/2} \le \varepsilon.
$$
where $\varepsilon$ is an appropriately chosen positive convergence tolerance value.

\vspace{.25cm}
\noindent\textbf{Ending phase of the PLS algorithm}

\vspace{.125cm}
\noindent Carry out the ordinary least squares estimation of the $\beta_{jk}$ coefficients linking $Y_k$ to $Y_j$ (for every inner submodel), the $\lambda_{jh}$ parameters (outer models, Mode A), specifying the linear relations between the latent $Y_j$ and the corresponding manifest $X_{jh}$ and the residual variances (having standardized the involved variables).

\section{An equivalent formulation of the PLS algorithm and its implementation with ordinal variables}
\label{AnequivalentreformulationofthePLSalgorithmanditsimplementationwithordinalvariables}

We now rewrite the PLS algorithm, by making extensive use of linear algebra notations, in order to avoid the reconstruction, at each step, of the latent scores. The procedure will be based on the covariance matrix of the observed manifest variables $X_{jh}$ or the polychoric correlation matrix in case of manifest variables of the ordinal type.
Namely, in presence of ordinal indicators we substitute the categorical variables $X_{jh}$ with the underlying latent variables $X^*_{jh}$, see (\ref{ZBBCequation003}), that are standardized and thus centered. Note that the components of $\bm X^*$ are not observable, but in the algorithm we will only make use of variances and covariances defined on their linear transformations.
\\These variances and covariances can be derived as a function of $\mathbf\Sigma_{XX}$, the covariance matrix of the vector random variable containing all the $(p+q)$ manifest indicators $X_{jh}$, of the metric/interval type, or their counterparts $X^*_{jh}$ when the indicators are ordinal; in the latter case $\mathbf\Sigma_{XX}$ is the polychoric correlation matrix defined across the ordered categorical variables.

\vspace{.125cm}

\noindent\textit{Step 0.} The outer approximation for the generic variable $Y_j$ is formally defined, see (\ref{boaricantaluppi.equation100}), as
$$
\hat Y_{j}=\sum^{p_j}_{h=1} w_{jh}^{(0)} (X_{jh} - \bar x_{jh})=\sum^{p_j}_{h=1} w_{jh}^{(0)} \,_CX_{jh}\equiv\sum^{p_j}_{h=1} w_{jh}^{(0)} X^*_{jh}
$$
Later on we will omit the symbol, here $^{(0)}$, specifying the iteration step of the algorithm.
\\Relationship (\ref{boaricantaluppi.equation100}) may be written in matrix form as
\begin{equation}
\label{equation120}
\hat Y_{j}= \left[0,\dots, w_{j1}, \dots, w_{jp_j},0,\dots,0\right] \left[
\begin{array}{c}
_CX_{11} \\ 
\vdots \\ 
_CX_{1p_1} \\ 
\vdots \\ 
_CX_{j1} \\ 
\vdots \\ 
_CX_{jp_j} \\ 
\vdots \\ 
_CX_{(n+m)1} \\ 
\vdots \\ 
_CX_{(n+m)p_{n+m}}
\end{array}
\right]
=\mathbf w'_j\,_C\bm X
\end{equation}
where $_C\bm X$ are the centered manifest variables that, in presence of ordinal indicators, are set to $_C\bm X\equiv\bm X^*$.
\\It is now possible to define the $N\times (m+n)$ matrix $\hat\mathbf Y= \left[\hat\mathbf Y_1,\dots,\hat\mathbf Y_{m+n}\right]$ containing the outer approximation values of the latent variables for the $N$ subjects as
$$
\hat\mathbf{Y}=\,_C\mathbf{XW}=\,_C\mathbf{X}
\left[
\mathbf{w}_1, \dots, \mathbf{w}_{j}, \dots, \mathbf{w}_{n+m}
\right],
$$
being $_C\mathbf X$ the $N\times(p+q)$ matrix of the deviations of the manifest variables from their means, and $\mathbf W=\left[\mathbf w_1, \dots, \mathbf w_{j}, \dots, \mathbf w_{n+m}\right]$ the square matrix containing the vectors $\mathbf w_j$ as columns.
\\Thus the covariance (\ref{equation135}) between $\hat Y_j$ and $\hat Y_k$ can be expressed as
\begin{equation}
\label{equation150}
Cov\left( \hat Y_j,\hat Y_k\right) = \mathbf w'_j E\left(\,_C\bm X\,_C\bm X' \right)\mathbf w_k=\mathbf w'_j\bm\Sigma_{XX}\mathbf w_k,
\end{equation}
and the variance covariance matrix of the random vector $\left ( \hat Y_1,\dots,\hat Y_{n+m} \right )$ as
\begin{equation}
\label{equation155}
\bm\Sigma_{\hat Y \hat Y}=\mathbf W'\bm\Sigma_{XX}\mathbf W.
\end{equation}
The standardized version (\ref{latstand}) of $\hat \mathbf Y$ is
$$
\tilde \mathbf Y=\hat \mathbf Y\left(\bm\Sigma_{\hat Y\hat Y}*\mathbf I\right)^{-1/2}=(\,_C\mathbf X)\mathbf W\left\{\left[\mathbf W'\bm\Sigma_{XX}\mathbf W\right]*\mathbf I\right\}^{-1/2}=(\,_C\mathbf{X})\,_S\mathbf{W}
$$
where $*$ is the Hadamard element by element product, $\mathbf I$ the identity matrix, and
\begin{equation}
\label{Wtostandardize}
_S\mathbf{W}= \mathbf W\left\{\left[\mathbf W'\bm\Sigma_{XX}\mathbf W\right]*\mathbf I\right\}^{-1/2}
\end{equation}
is a transformation of the original weights $\mathbf W$, which for each group of manifest indicators sum up to 1, into a set of weights allowing the latent variables to be on a standardized scale. We then set
$$
\hat \mathbf Y=\tilde \mathbf Y.
$$
The covariance matrix across standardized $\left (\hat Y_1,\dots,\hat Y_{n+m} \right )$ can be re-defined as
\begin{equation}
\label{covacorlat}
\bm\Sigma_{\hat Y \hat Y}=\mathbf P_{\hat Y \hat Y}=\,_S\mathbf{W}'\bm\Sigma_{XX}\,_S\mathbf{W}
\end{equation}
so becoming a correlation matrix.

\vspace{.25cm}
\noindent\textbf{Iterative phases of the PLS algorithm}

\vspace{.125cm}
\noindent\textit{Step 1.} 
The instrumental variables $Z_j$, see (\ref{equation130}), which are defined for each latent variable $Y_j$ as a linear combination of the estimates of the latent variables $Y_k$ linked to $Y_j$ in the path diagram may be expressed as
\begin{eqnarray}
Z_j&=& \bm\tau'_j
\left[
\begin{array}{c}
\hat Y_1 \\
\vdots \\
\hat Y_{n+m}
\end{array}
\right]=
\left[0,\dots, 0,\dots,1,\dots,0,1,0\right] \left[
\begin{array}{c}
_S\mathbf{w}'_1 \\ 
\vdots \\ 
_S\mathbf{w}'_{n+m}
\end{array}
\right]
\left[
\begin{array}{c}
_CX_{11} \\ 
\vdots \\ 
_CX_{1p_1} \\ 
\vdots \\ 
_CX_{j1} \\ 
\vdots \\ 
_CX_{jp_j} \\ 
\vdots \\ 
_CX_{(n+m)1} \\ 
\vdots \\ 
_CX_{(n+m)p_{n+m}}
\end{array}
\right]= \nonumber \\
\label{equation140}
&=&\bm\tau'_j\left[
\begin{array}{c}
_S\mathbf{w}'_1 \\ 
\vdots \\ 
_S\mathbf{w}'_{n+m}
\end{array}
\right]\,_C\bm X
=
\bm\tau'_j
\,_S\mathbf{W}'
\,_C\bm X
\end{eqnarray}
where $\bm\tau_j$ is the $j$th column of the matrix $\bm\Upsilon$ with generic element $\tau_{jk}$ defined, according to (\ref{equation130}), as
\begin{equation}
\label{equation156}
\bm\Upsilon=(\mathbf T+\mathbf T')*sign(\bm\Sigma_{\hat Y\hat Y})
\end{equation}
being the elements of $\mathbf T$ equal to 0 or 1.
\\With reference to the matrix $_C\mathbf{X}$ of observable data the $N\times (m+n)$ matrix $\mathbf Z$ containing the values of the $(n+m)$ instrumental variables $Z_1,\dots,Z_{n+m}$ for the $N$ subjects may be obtained as
$$
\mathbf Z=[\mathbf Z_1,\dots,\mathbf Z_j,\dots,\mathbf Z_{n+m}]=\,_C\mathbf{X}\,_S\mathbf{W}\bm\Upsilon.
$$

\vspace{.125cm}
\noindent\textit{Step 2.} The covariance (\ref{boaricantaluppi.equation16014}) between $X_{jh}$ and $Z_j$ is defined as
\begin{eqnarray}
Cov\left( X_{jh},Z_j\right) &=& Cov\left(X_{jh},
\bm\tau'_j
\left[
\begin{array}{c}
\hat Y_1 \\
\vdots \\
\hat Y_{n+m}
\end{array}
\right]
\right) = \nonumber \\
&=& Cov\left(X_{jh},
\bm\tau'_j
\left[
\begin{array}{c}
_S\mathbf{w}'_1 \\ 
\vdots \\ 
_S\mathbf{w}'_{n+m}
\end{array}
\right]
\left[
\begin{array}{c}
_CX_{11} \\ 
\vdots \\ 
_CX_{1p_1} \\ 
\vdots \\ 
_CX_{j1} \\ 
\vdots \\ 
_CX_{jp_j} \\ 
\vdots \\ 
_CX_{(n+m)1} \\ 
\vdots \\ 
_CX_{(n+m)p_{n+m}}
\end{array}
\right]\right)= \nonumber \\
\label{equation160}
&=&\bm\tau'_j
\left[
\begin{array}{c}
_S\mathbf{w}'_1 \\ 
\vdots \\ 
_S\mathbf{w}'_{n+m}
\end{array}
\right]
\left[
\begin{array}{c}
\sigma_{X_{11}X_{jh}} \\ 
\vdots \\ 
\sigma_{X_{1p_1}X_{jh}} \\ 
\vdots \\ 
\sigma_{X_{j1}X_{jh}} \\ 
\vdots \\ 
\sigma_{X_{jp_j}X_{jh}} \\ 
\vdots \\ 
\sigma_{X_{(n+m)1}X_{jh}} \\ 
\vdots \\ 
\sigma_{X_{(n+m)p_{n+m}}X_{jh}} \\ 
\end{array}
\right]
\end{eqnarray}
and $\bm\Sigma_{XZ}$ the covariance matrix between all the manifest variables $X_{jh}$ and the instrumental variables $Z_j$ as
\begin{equation}
\label{equation165}
\bm\Sigma_{XZ}=\bm\Sigma_{XX}\,_S\mathbf {W}\bm\Upsilon.
\end{equation}
The covariance between $X_{jh}$ and $\hat Y_j$ is
\begin{equation}
\label{equation170}
Cov\left( X_{jh},\hat Y_j\right) = Cov\left(X_{jh},
\,_S\mathbf{w}'_j
\left[
\begin{array}{c}
_CX_{j1} \\ 
\vdots \\ 
_CX_{jp_j} \\ 
\end{array}
\right]\right) =
\,_S\mathbf{w}'_j
\left[
\begin{array}{c}
\sigma_{X_{j1}X_{jh}} \\ 
\vdots \\ 
\sigma_{X_{jp_j}X_{jh}} \\ 
\end{array}
\right]
\end{equation}
and $\bm\Sigma_{XY}$ the covariance matrix between all the manifest variables $X_{jh}$ and the composites $\hat Y_j$ can be obtained as
\begin{equation}
\label{equation175}
\bm\Sigma_{X\hat Y}=\bm\Sigma_{XX}\,_S\mathbf{W}.
\end{equation}
Now define a $(p+q)\times(n+m)$ matrix $\mathbf C$ by the Hadamard product of the indicator matrix $\bm\chi_{\mathbf W}$ of the matrix $\mathbf W$ and the covariance matrix between $X$ and $Z$
\begin{equation}
\label{equation176}
\mathbf C=\bm\chi_{\mathbf W}*\bm\Sigma_{XZ};
\end{equation}
it results in a block diagonal matrix with generic block
$$
[\mathbf C_{jj}]=
\left[
\begin{array}{c}
\sigma_{X_{j1}Z_j} \\ \vdots \\ \sigma_{X_{jp_j}Z_j} \\
\end{array}
\right], \quad j=1,\dots,m+n.
$$
The matrix $\mathbf W$ with the updated weights is obtained from
\begin{equation}
\label{equation177}
\mathbf W=\mathbf C[diag(\mathbf 1'_{p+q}\mathbf C)]^{-1}diag(\pm),
\end{equation}
where $\mathbf 1$ is the $(p+q)\times 1$ unitary vector, $diag(\cdot)$ is the operator transforming a vector in a diagonal matrix and $\pm$ is the vector defined as
\begin{equation}
\label{equation178}
\pm=sign \left\{\mathbf 1'_{p+q}\left[sign\left(\bm\chi_{\mathbf W}\bm\Sigma_{X\hat Y}\right)\right]\right\}.
\end{equation}
Finally transformation (\ref{Wtostandardize}) may be applied to obtain the standardizing weights $_S\mathbf{W}$.

We resume the sequence of steps defining the reformulation of the PLS algorithm, which has the characteristic of avoiding the determination, at each step, of the composites scores $\hat y_js$ and of the instrumental variables scores $z_js$. Ending phases of the PLS algorithm will be described later (see Sections \ref{EndingphaseofthePLSalgorithmwhenmanifestvariablesareoftheintervaltype} and \ref{ParameterEstimationoftheinnerandouterrelationshipsinpresenceofordinalmanifestvariables}).

\vspace{.2cm}
  Compute $\bm\Sigma_{XX}$ (in case of ordinal items the polychoric correlation matrix)

\vspace{.2cm}
  Define the starting weights $\mathbf W=
  \left[
  \mathbf w_1, \dots, \mathbf w_{j}, \dots, \mathbf w_{n+m}
  \right].
  $

\vspace{.2cm}
\textit{Iterative phase}

\vspace{.2cm}
\rule{1.04cm}{0cm} Set
$
\mathbf W_{\mathrm{TEMP}}=\mathbf W
$

\vspace{.2cm}
\rule{.52cm}{0cm} Compute:
  
\vspace{.2cm}
\rule{1.04cm}{0cm} $
\bm\Sigma_{\hat Y \hat Y}=\mathbf W'\bm\Sigma_{XX}\mathbf W \quad \quad \mathrm{see\ (\ref{equation155})}
$

\vspace{.2cm}
  
\rule{1.04cm}{0cm} $
_S\mathbf{W}= \mathbf W\left\{\left[\mathbf W'\bm\Sigma_{XX}\mathbf W\right]*\mathbf I\right\}^{-1/2}= \mathbf W\left[\Sigma_{\hat Y \hat Y}*\mathbf I\right]^{-1/2} \quad \quad \mathrm{see\ (\ref{Wtostandardize})}
$

\vspace{.2cm}
  
\rule{1.04cm}{0cm} $
\bm\Sigma_{\hat Y \hat Y}=\mathbf P_{\hat Y \hat Y}=\,_S\mathbf{W}'\bm\Sigma_{XX}\,_S\mathbf{W} \quad \quad \mathrm{see\ (\ref{covacorlat})}
$

\vspace{.2cm}
  
\rule{1.04cm}{0cm} $
\bm\Upsilon=(\mathbf T+\mathbf T')*sign(\bm\Sigma_{\hat Y\hat Y}) \quad \quad \mathrm{see\ (\ref{equation156})}
$

\vspace{.2cm}
  
\rule{1.04cm}{0cm} $
\bm\Sigma_{XZ}=\bm\Sigma_{XX}\,_S\mathbf{W}\bm\Upsilon \quad \quad \mathrm{see\ (\ref{equation165})}
$

\vspace{.2cm}
  
\rule{1.04cm}{0cm} $
\bm\Sigma_{X\hat Y}=\bm\Sigma_{XX}\,_S\mathbf{W} \quad \quad \mathrm{see\ (\ref{equation175})}
$

\vspace{.2cm}
  
\rule{1.04cm}{0cm} $
\mathbf C=\bm\chi_{\mathbf W}*\bm\Sigma_{XZ} \quad \quad \mathrm{see\ (\ref{equation176})}
$

\vspace{.2cm}
  
\rule{1.04cm}{0cm} $
\pm=sign \left\{\mathbf 1'_{p+q}\left[sign\left(\bm\chi_{\mathbf W}\bm\Sigma_{X\hat Y}\right)\right]\right\} \quad \quad \mathrm{see\ (\ref{equation178})}
$

\vspace{.2cm}
\rule{.52cm}{0cm} Update the weights $
\mathbf W=\mathbf C[diag(\mathbf 1'_{p+q}\mathbf C)]^{-1}diag(\pm) \quad \quad \mathrm{see\ (\ref{equation177})}
$

\vspace{.2cm}
\rule{.52cm}{0cm} Obtain $_S\mathbf{W} \quad \quad \mathrm{see\ (\ref{Wtostandardize})}$

\vspace{.2cm}
  Check if 
$
||\mathbf W-\mathbf W_{\mathrm{TEMP}}||<\varepsilon
$

\subsection{Ending phase of the PLS algorithm with manifest variables of the interval type}
\label{EndingphaseofthePLSalgorithmwhenmanifestvariablesareoftheintervaltype}
After convergence of the weights in $\mathbf W$ the score values can be determined as
$$
\hat\mathbf Y=\,_C\mathbf{X}\,_S\mathbf{W}
$$
and OLS regressions carried out (on standardized variables) to obtain the estimates of the parameters in $\mathbf D$ and $\bm\Lambda$ and the variances of the error components as the residual variances of the corresponding regression models.

\subsection{Parameter Estimation of the inner and outer relationships in presence of ordinal manifest variables}
\label{ParameterEstimationoftheinnerandouterrelationshipsinpresenceofordinalmanifestvariables}

The estimates in the inner and outer regression models can also be obtained without having to reconstruct the score values $\hat y_{js}$, that cannot be estimated in presence of manifest variables of the ordinal type: their prediction may be obtained according to one of the procedures illustrated in Section \ref{Prediction}.

The parameter estimates make reference to the following linear regression models, defined on standardized variables,
\begin{equation}
\label{equationinnerlinearmodel}
\hat Y_j=\sum_{i=1}^d \,_j\beta_i\,_{Rj}\hat Y_i+\nu_j,\quad j=m+1,\dots,m+n
\end{equation}
where $d<j$ and $\{\,_{Rj}\hat Y_1,\dots,\,_{Rj}\hat Y_d\}$ is a subset of $\{\hat Y_1,\dots,\hat Y_{j-1}\}$ defined according to the $j$th equation in (\ref{equation030}).

The estimate of the vector $\,_j\bm\beta=[\,_j\beta_1,\dots,\,_j\beta_d]'$, which contains the unknown elements of the $m+j$ row of matrix $\mathbf D$ in (\ref{equation030}), may be computed as
\begin{equation}
\label{equationinnerlinearmodelest}
_j\hat{\bm\beta}=\,_{Rj}\bm\Sigma_{\hat Y\hat Y}^{-1}\,_j\bm\Sigma_{\hat Y\hat Y}
\end{equation}
where $_{Rj}\bm\Sigma_{\hat Y\hat Y}$ is the matrix obtained by extracting from $\bm\Sigma_{\hat Y\hat Y}=\mathbf P_{\hat Y\hat Y}$, the correlation matrix of $\hat\mathbf Y$, see (\ref{covacorlat}), the rows and columns pertaining to the independent variables in the linear model (\ref{equationinnerlinearmodel}) and  $_j\bm\Sigma_{\hat Y\hat Y}$ is the vector obtained by extracting from the $j$th column of $\bm\Sigma_{\hat Y\hat Y}$ the elements corresponding to correlations between $\hat Y_j$ and its covariates, according to relationship (\ref{equationinnerlinearmodel}).

Let now $\bm\Sigma_{X\hat Y}$ be the correlation matrix between the manifest indicators $\bm X$ and the composites $\hat\mathbf Y$, which can be derived from (\ref{equation175}).
The estimate of parameter $\lambda_{jh}$ in the outer model
$$
X_{jh}=\lambda_{jh}\hat Y_j+\varepsilon_{jh},\quad j=1,\dots,m+n, h=1,\dots,p_j
$$
is given by the correlation coefficient between $X_{jh}$ and $\hat Y_j$.

The ending phase of the PLS algorithm, in presence of manifest variables of the ordinal type, can be then resumed in the following way:

\vspace{.2cm}
\rule{.52cm}{0cm} $_j\hat{\bm\beta}=\,_{Rj}\mathbf P_{\hat Y\hat Y}^{-1}\,_j\mathbf P_{\hat Y\hat Y}$

\vspace{.2cm}
\rule{.52cm}{0cm} $\hat\lambda_{jh}$ is equal to the correlation coefficient between $X_{jh}$ and $\hat Y_j$.

\vspace{.25cm}
\noindent In presence of manifest variables of the interval type the procedure gives, by having recourse to Pearson's correlations, the same results as the ending phase of the usual PLS algorithm presented in Section \ref{EndingphaseofthePLSalgorithmwhenmanifestvariablesareoftheintervaltype}.

Only the 'covariance' or 'correlation' matrix of the manifest variables $\bm X$ is needed in order to determine the final weights and the inner and outer model parameter estimates. In presence of manifest indicators of the ordinal type we propose to use the polychoric correlation matrix. This is consistent with the so-called METRIC 1 option suggested by \cite{Lohmoeller_1989} \citep[see also][]{Tenenhaus2005,Rigdon2012} performing the standardization of all manifest indicators.

Both polychoric covariance and correlation matrices must be invertible for the above procedure to work.
However, constrained algorithms do exist whenever invertibility problems should arise for the polychoric correlation matrix. (For example the function \textit{hetcor}, available in the R \textit{polycor} package, makes it possible to require that the computed matrix of pairwise polychoric, polyserial, and Pearson correlations be positive-definite).

After having transformed the manifest variables according to (\ref{ZBBCequation003}) the threshold values related to the standard normal variables $X^*_1,\dots,X^*_{m+n}$ are available.
In this case we have $\bm\Sigma_{X^*X^*}\equiv\mathbf P_{X^*X^*}$, that is the polychoric covariance matrix between the underlying latent variables coincides with the polychoric correlation matrix.

\vspace{.25cm}
The algorithm we have presented for ordinal manifest variables will be denoted as Ordinal Partial Least Squares (OPLS) from now on.

\subsection{Prediction of Latent Scores}
\label{Prediction}

A point estimate of the composite continuous (latent) $Y_j$ cannot be determined in presence of ordinal variables for the generic subject; we can only establish an interval of possible values conditional on the threshold values pertaining the latent variables $X^*_{jh}$ that underlie each ordinal manifest variable.

Since each underlying $X^*_{jh}$ variable is assumed to be a standard Normal variate, the composite variable $\hat Y_j$, defined by the outer approximation
$$
\hat Y_j=\sum^{p_j}_{h=1} \,_Sw_{jh} X^*_{jh},
$$
will also be distributed according to a standard Normal variate.

A set of threshold values $a^{Y_j}_i$, $i=1,\dots,I-1$, can be derived from the threshold values $a^{X_{jh}}_i$ referred to the variables $X^*_{jh}$, $h=1,\dots,p_j$, being $I$ the common number of categories assumed by the variables $X_{jh}$, as
$$
a^{Y_j}_i=\sum^{p_j}_{h=1} \,_Sw_{jh} \, a^{X_{jh}}_i.
$$
Should some threshold values equal $\pm\infty$ they have to be replaced with $\pm 4$. Later we will also consider $a^{Y_j}_0=-4$ and $a^{Y_j}_I=4$.

For the generic subject $s$ expressing the values $x_{jhs}$ for the variables $X_{jh}, h=1,\dots,p_j$, linked to the generic $Y_j$, let us first define the sets of $y$ values images of all possible $x_{jhs}$.

In case subject $s$ chooses the same category $i$ for all the manifest indicators of $Y_j$, that is $x_{j1s}=\dots=x_{jp_js}=i$ with $i \in \{1,\dots,I\}$, the image will be of the type
$$
A_i\equiv (a^{Y_j}_{i-1},a^{Y_j}_i]
$$
which we will call 'homogeneous thresholds'.

Otherwise, see Fig. \ref{latentthresholds}, the set which is the image of all possible responses $x_{jhs}$, will not correspond exactly to one subset $A_i$. Let us denote this set with
$$
C_{js}\equiv (\alpha^{Y_j}_s,\beta^{Y_j}_s]
$$
where:
$$
\alpha^{Y_j}_s=\sum^{p_j}_{h=1} \,_Sw_{jh} \, a^{x_{jhs}}_{i-1} \quad \quad \mathrm{and} \quad \quad \beta^{Y_j}_s=\sum^{p_j}_{h=1} \,_Sw_{jh} \, a^{x_{jhs}}_i,
$$
being $a^{x_{jhs}}_{i-1}$ and $a^{x_{jhs}}_i$ the threshold values corresponding to the category $x_{jhs}$ observed by subject $s$, that is the values defining the interval for $X^*_{jh}$ to have $x_{jhs}$ as image according to (\ref{ZBBCequation003}).
\\In order to assign a category to subject $s$ for the latent variable $Y_j$ we can use one of the following options:
\begin{enumerate}
  \item \textbf{Mode estimation.} Compute, see Fig. \ref{latentthresholds}, the probabilities for $C_{js}$ to overlap each set $A_i$ defined by the 'homogeneous thresholds'
  \begin{equation}
  \label{categoryassignmentmode}
  P\left(C_{js}\cap A_i\right) \quad i=1,\dots,I
  \end{equation}
  and, for subject $s$, select the set $A_i$ with maximum probability.
  To the set $A_i$ corresponds the assignment of category $i$ as a score estimate for the latent variable $Y_j$.
  \begin{figure}[t]
  \begin{center}
  \includegraphics[scale=.6]{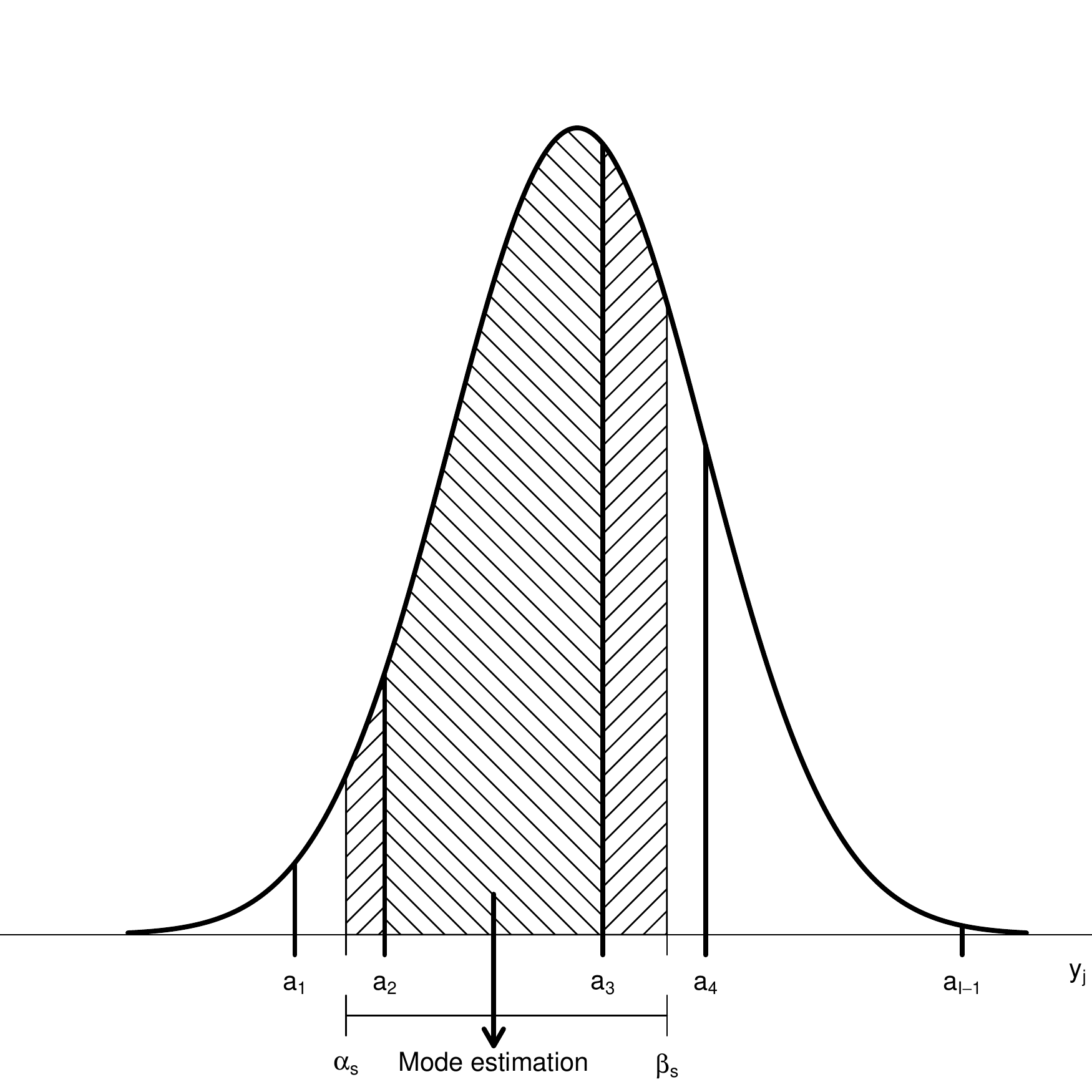}
  \end{center}
  \caption{Latent variable category assignment, see (\ref{categoryassignmentmode})}
  \label{latentthresholds}
  \end{figure}
  \item \textbf{Median estimation.} Compute the median of the variable $Y_j$ over the interval $C_{js}$
  \begin{equation}
  \label{categoryassignmentmedian}
  Median(Y_j|Y_j \in C_{js}) = \Phi^{-1}\left (\frac{1}{2}(\Phi(\alpha^{Y_j}_s) + \Phi(\beta^{Y_j}_s) \right )
  \end{equation}
  the category $i$ pertaining the set $A_i$ to which $Median(Y_j|Y_j \in C_{js})$ belongs, is assigned to subject $s$.
  \item \textbf{Mean estimation.} Compute the mean of the variable $Y_j$ over the interval $C_{js}$
  \begin{equation}
  \label{categoryassignmentmean}
  E(Y_j|Y_j \in C_{js})=\frac{\phi(\alpha^{Y_j}_s)-\phi(\beta^{Y_j}_s)}{\Phi(\beta^{Y_j}_s)-\Phi(\alpha^{Y_j}_s)}
  \end{equation}
  the category $i$ pertaining the set $A_i$ to which $E(Y_j|Y_j \in C_{js})$ belongs, is assigned to subject $s$.
\end{enumerate}

\subsection{Bias effects of the OPLS algorithm}
\label{EffectsofthebiasofthePLSalgorithmoncategoricalPLS}
\cite{Schneeweiss_1993} shows that parameter estimates obtained with the PLS algorithm are negatively biased for the inner model, (these estimates are related to the covariances or correlations across latent variables). The OPLS is based on the analysis of the polychoric correlation matrix, which is obtained by maximizing the correlation across the latent variables that generate, according to (\ref{ZBBCequation003}), the manifest variables. Especially in presence of items with a low number of categories, polychoric correlations are usually larger than the Pearson's ones as was also observed by \cite{Coenders1997}. Thus we may expect that the distribution of the inner model parameter estimates obtained with OPLS dominates stochastically that of the PLS algorithm, possibly reducing the negative bias of estimates based on Pearson's correlations.

However, the reduction in the bias of the inner model parameter estimates for OPLS can have a drawback: a positive bias in the estimation of the parameters in the outer model.
\cite{FornellCha_1994} report, for the special case of identical correlation coefficients (say $\rho$) across all the manifest variables, the following relationship relating the bias of the PLS algorithm with respect to maximum likelihood estimates of the outer model $\lambda$ parameters and the one referred to the common correlation across latent variables, upon which the inner model coefficients $\beta$ are obtained:
$$
\mathrm{bias}(\lambda)=\frac{1}{\sqrt{\mathrm{bias}(\rho)}}.
$$
An high value for bias$(\lambda)$ corresponds to a low value of bias$(\rho)$; we have observed this issue in the illustration presented in Section \ref{SomeSimulations}.

Anyway, we have to observe that outer model parameters are not the most important target in a decision making procedure based on the PLS estimation of a structural equation model with latent variables: the main role is played by the inner model parameter estimates and by the weights $w_{jh}$ see (\ref{boaricantaluppi.equation100}) and (\ref{equation120}), defining each PLS latent variable as a 'composite', that is a linear combination of its related manifest variables. The largest weights are related to the manifest variables which are supposed to have greater influence in driving the 'composite'; moreover, since the weights sum up to 1 they should not suffer of any dimensional bias problem.

\subsection{Assessing reliability}
Scale reliability can be assessed, for ordinal scales, by having recourse to methods based on the polychoric correlation matrix \citep[see][]{Zumbo2007,Zumbo2012} for Cronbach's $\alpha$. The Dillon-Goldstein's rho \citep{Chin_1998} and methods presented for covariance based models \citep{GreenYang2009,Raykov}, which make reference to all relationships in the structural equation model (\ref{equation010})-(\ref{equation045}), can also be implemented.

\section{Illustrative examples}
\label{Applications}

The Partial Least Squares algorithm has been successfully applied to estimate models aimed at measuring customer satisfaction, first at a national level \citep{FornellSwedish_1996,Fornell_1996} and then also in a business context \citep{Johnson_2000,Johnson_2001}. A widespread literature on this field is available.

The OPLS methodology is here implemented in R \citep{citeRcite} and applied to a well-known data set describing the measure of customer satisfaction in the mobile phone industry \citep{Bayol_2000,Tenenhaus2005}. 
By means of this example we compare the behaviour of the PLS and OPLS in presence of a traditional questionnaire whose items are characterized by a high number of categories (say 10).

Some simulations are then reported to analyze the behaviour of the procedure when the number of points for each item is reduced.

The R procedures by \cite{Fox_2010} and \cite{Revelle_2011} are used to compute polychoric correlation matrices, with minor changes to allow polychoric correlations to be computed when the number of categories is larger than 8. We never needed the polychoric correlation matrix to be forced in order to comply with the positive definiteness condition.

\subsection{The Mobile Phone Data Set}
\label{TheMobilePhoneDataSet}

We applied the procedure to a classical example on mobile phone, presented in \cite{Bayol_2000} and \cite{Tenenhaus2005}.
Data (250 observations) are available e.g. in \cite{Trinchera_2010}. Data were collected on 24 ordered categorical variables with 10 categories; the observed variables are resumed by 7 latent variables.

\begin{figure}
\begin{center}
\includegraphics[scale=1]{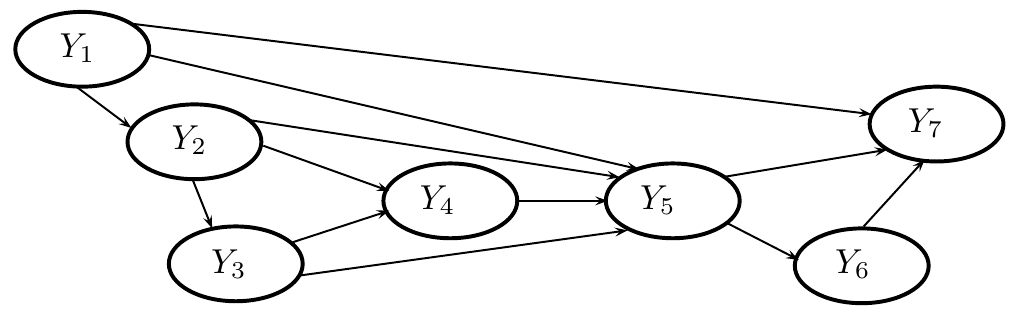}
\end{center}
\caption{Path diagram for the mobile phone industry customer satisfaction model}
\label{mobilephonefig}
\end{figure}

The customer satisfaction model underlying the mobile phone data refers to a version of the European Customer Satisfaction Index with 1 exogenous latent variable, the Image ($Y_1$) with 5 manifest indicators, and 6 endogenous latent variables: Customer Expectations ($Y_2$), Perceived Quality ($Y_3$), Perceived Value ($Y_4$), Customer Satisfaction ($Y_5$), Loyalty ($Y_6$) and Complaints ($Y_7$), with respectively $3,7,2,3,1$ and 3 manifest indicators.
See Fig. \ref{mobilephonefig} for the inner path model relationships.
Table 1 in \cite{Tenenhaus2005} contains the structure of the questionnaire; it can be considered as a possible instrument for customer satisfaction measurement in the mobile phone industry.

Table \ref{Bayolbetaparcomp} reports the parameter estimates obtained both with the standard PLS algorithm and with the OPLS algorithm.
\begin{table}
\caption{Mobile phone industry customer satisfaction model: PLS and OPLS parameter estimates with their significance  (s.e. in brackets)}
\small
\begin{center}
\begin{tabular}{cccccc}
  \hline
 & \multicolumn{2}{c}{PLS}  & &  \multicolumn{2}{c}{OPLS}   \\ 
  \hline
  $\beta_{21}$ & 0.491 & (0.000) & & 0.584 & (0.000) \\ 
  $\beta_{32}$ & 0.545 & (0.000) & & 0.612 & (0.000) \\ 
  $\beta_{42}$ & 0.067 & (0.281) & & 0.037 & (0.563) \\ 
  $\beta_{43}$ & 0.540 & (0.000) & & 0.596 & (0.000) \\ 
  $\beta_{51}$ & 0.153 & (0.006) & & 0.199 & (0.001) \\ 
  $\beta_{52}$ & 0.035 & (0.431) & & 0.035 & (0.423) \\ 
  $\beta_{53}$ & 0.544 & (0.000) & & 0.517 & (0.000) \\ 
  $\beta_{54}$ & 0.201 & (0.000) & & 0.198 & (0.000) \\ 
  $\beta_{65}$ & 0.541 & (0.000) & & 0.563 & (0.000) \\ 
  $\beta_{71}$ & 0.212 & (0.001) & & 0.261 & (0.000) \\ 
  $\beta_{75}$ & 0.466 & (0.000) & & 0.493 & (0.000) \\ 
  $\beta_{76}$ & 0.051 & (0.376) & & 0.043 & (0.417) \\ 
   \hline
\end{tabular}
\end{center}
\label{Bayolbetaparcomp}
\end{table}

Surprisingly, results are quite similar but not so close. When the number of categories is sufficiently high, Pearson correlation coefficients are good approximations for their polychoric counterparts, but responses in customer satisfaction surveys do usually have skewed distributions since respondents do not effectively choose, with a non-negligible frequency, all the available categories of the manifest variables. 6 over 24 manifest indicators had only 6 categories with at least 5 respondents. Thus differences between Pearson and polychoric correlations may be evident with effects on parameter estimates.
\\Coefficients computed with the OPLS algorithm, that are significantly different from 0, are larger (except for $\beta_{53}$ and $\beta_{54}$) than those obtained with the PLS algorithm which is known to underestimate the inner model parameters \citep[see][]{Schneeweiss_1993} and it is also based on Pearson's correlations which underestimate real correlations when the ordinal manifest variables are measured on scales with a small number of categories.

\begin{figure}
\begin{center}
\includegraphics[scale=.45]{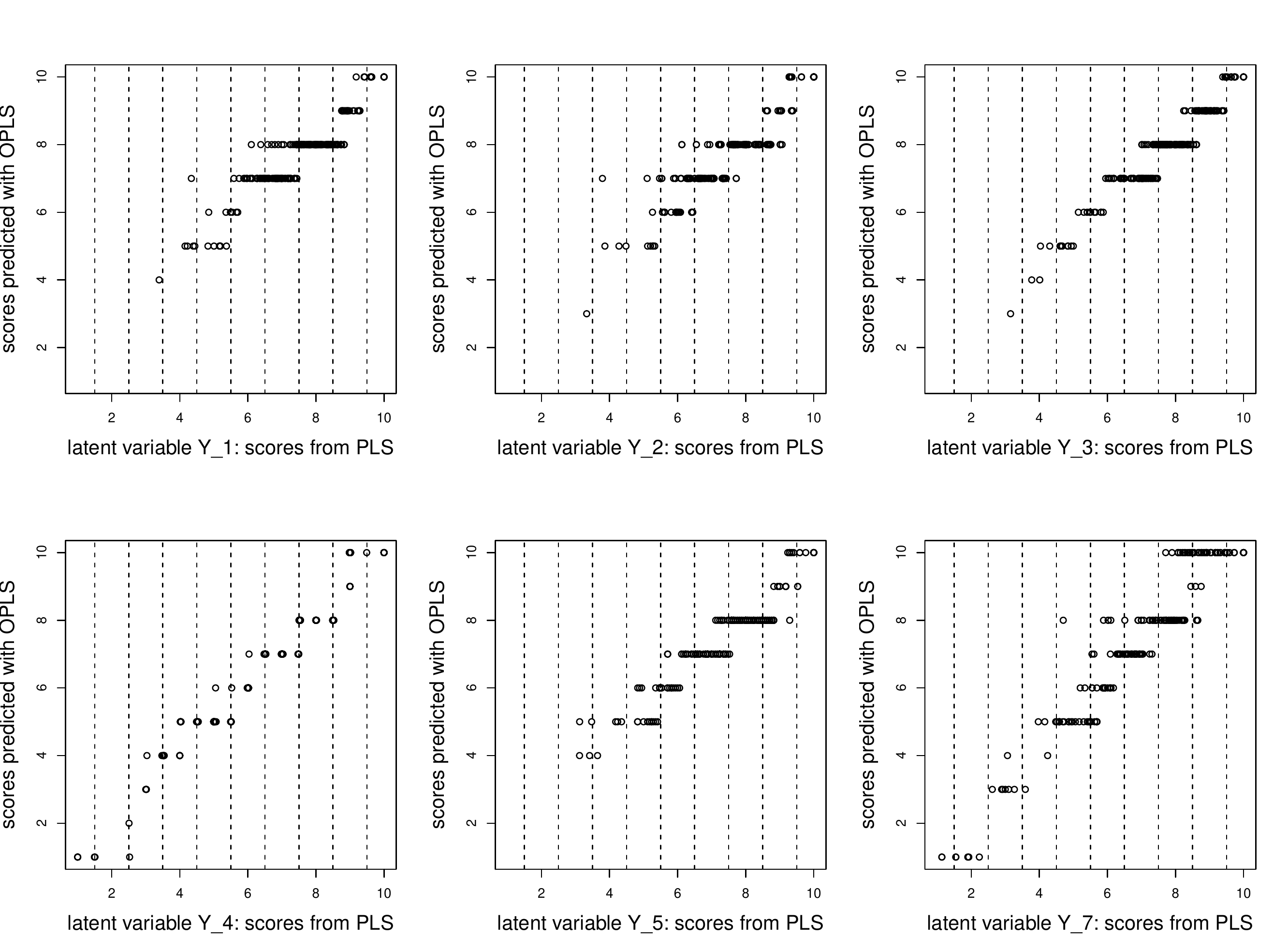}
\end{center}
\caption{PLS and OPLS (Mode estimation) latent variable score comparison}
\label{scorecomparison}
\end{figure}

Figure \ref{scorecomparison} shows a comparison of the latent scores reconstructed with the two methodologies. Information about $Y_6$ is not reported since the variable is identical to its unique manifest indicator.
Recall that according to the PLS algorithm the scores are weighted averages of the values expressed by the subjects on the proxies; the scores are thus generated on an interval scale. With the OPLS algorithm latent variable scores can only be predicted according to one of the procedures presented in Section \ref{Prediction} and their values are on the same ordinal scale common to the proxy variables (the procedure 'Mode estimation' was adopted to produce the graph).
\\Table \ref{Coherencyoflatentscores} shows the degree of coherency of the latent scores obtained with the traditional PLS algorithm and the 3 procedures presented in Section \ref{Prediction} for OPLS.
Having rounded scores obtained with the PLS algorithm to integer values, percentages of exact concordance are reported on the first three lines, while in the remaining lines are percentages of concordance with a difference between rounded values not larger than 1. We have at least 70\% exact concordance (except for latent variable $Y_7$); more than 90\% of cases for all latent variables show a difference between rounded values lower than 1.
\begin{table}
\caption{Coherency of latent scores between PLS and OPLS}
\small
\begin{center}
\begin{tabular}{lcccccc}
\hline
Method & $Y_1$ & $Y_2$ & $Y_3$ & $Y_4$ & $Y_5$ & $Y_7$ \\
\hline
Mode estimation$^a$   & 70.0 & 71.6 &  79.2 &  84.4 & 71.6 & 49.2 \\
Median estimation$^a$ & 74.8 & 75.2 &  78.0 &  88.0 & 70.4 & 51.2 \\
\vspace{.125cm} 
Mean estimation$^a$   & 72.8 & 77.2 &  75.6 &  86.8 & 71.6 & 50.4 \\
Mode estimation$^b$   & 98.8 & 98.0 & 100.0 &  99.6 & 99.2 & 89.6 \\
Median estimation$^b$ & 99.2 & 98.4 & 100.0 & 100.0 & 99.2 & 94.0 \\
Mean estimation$^b$   & 99.2 & 98.4 & 100.0 &  99.6 & 99.6 & 90.0 \\
\hline
\end{tabular}
\\$^a$ percentages of exact concordance after having rounded PLS scores to integer values
\\$^b$ percentages of concordance with a difference between rounded values not larger than 1
\end{center}
\label{Coherencyoflatentscores}
\end{table}

\begin{figure}
\begin{center}
\includegraphics[scale=.46]{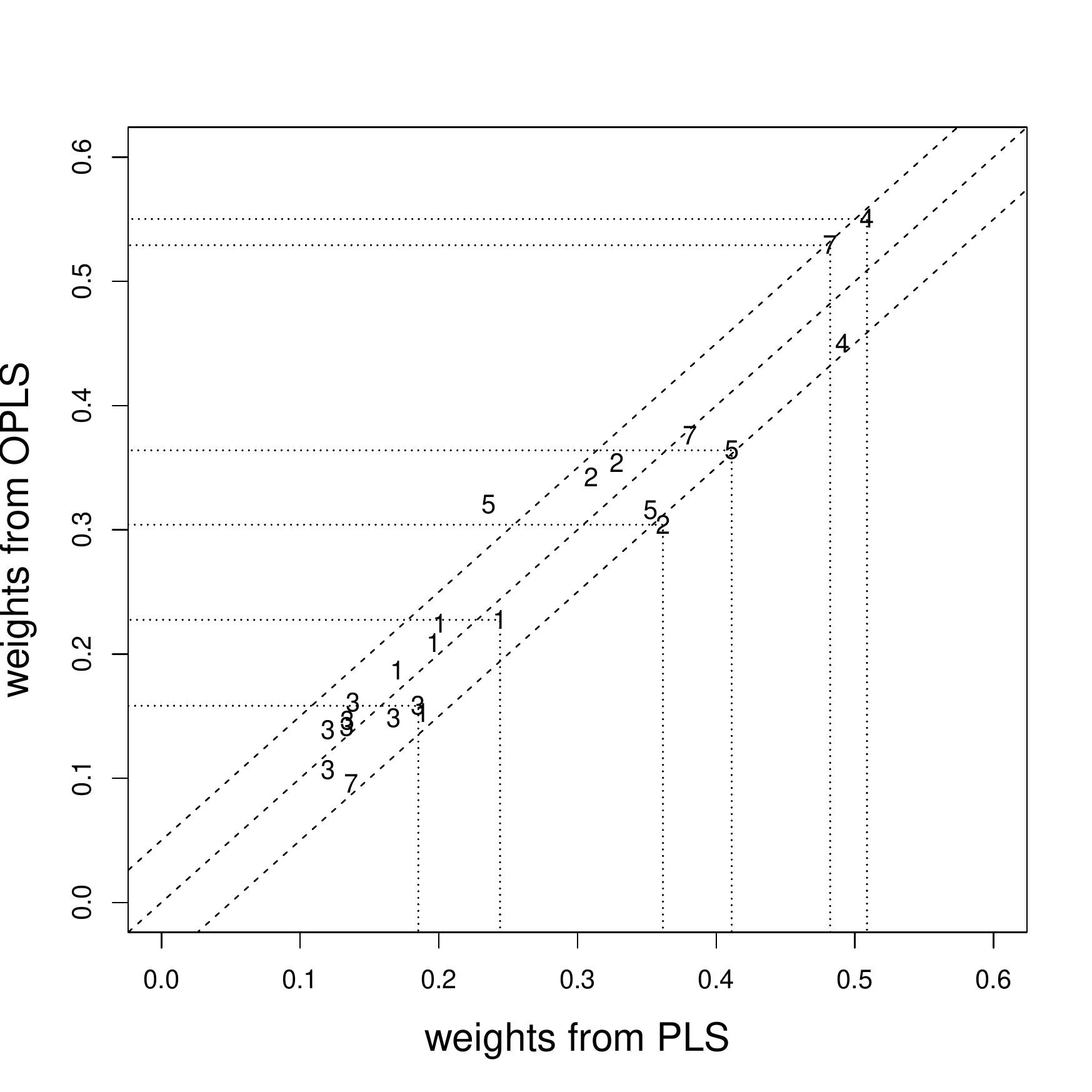}
\end{center}
\caption{PLS and OPLS weights comparison}
\label{weightscomparison}
\end{figure}

The weights $w_j$, see relationships (\ref{boaricantaluppi.equation16013}) or (\ref{equation177}), play an important role in making decisions based on PLS estimation of structural equation model with latent variables. They establish which proxy variables drive a latent variable behaviour. Latent variables are defined as 'composite' variables in the PLS algorithm, that is weighted averages of their manifest indicators with weights $w_j$. Figure \ref{weightscomparison} shows the comparison of the weights for the 7 latent variables obtained with the two algorithms. Points with the same number identify the weights assigned by the two algorithms to the manifest indicators of each latent variable $Y_j, j=1,2,\ldots,5,7$ (values 6 do not appear since $Y_6$ has only one indicator with unitary weight).
Dashed bandwidths include pairs which differ, with the two methodologies, no more than 0.05.
Only 2 values over the 23 weights are outside the bandwidths.
We can conclude that the two methods construct composite variables in quite the same manner. Dotted lines give information for each latent variable about the indicator with the largest weight determined by the two algorithms: except for latent variables $Y_2$ and $Y_3$ the same proxy variable is identified as the most important in explaining each latent construct.

\subsection{Some Simulations}
\label{SomeSimulations}
\begin{figure}
\begin{center}
\includegraphics[scale=1]{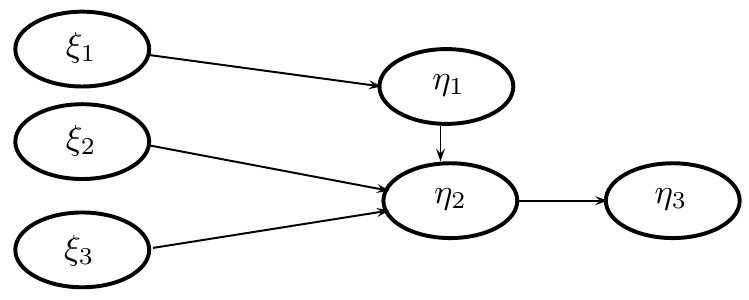}
\end{center}
\caption{Path diagram for simulated models}
\label{pathfig}
\end{figure}

To compare the performance of the classical PLS algorithm with the OPLS for different number of points in the scale of manifest variables we considered some simulations from the following model
$$
\begin{array}{l}
\eta_1=\gamma_{11}\xi_1+\zeta_1 \\
\eta_2=\beta_{21}\eta_1+\gamma_{22}\xi_2+\gamma_{23}\xi_3+\zeta_2 \\
\eta_3=\beta_{32}\eta_2+\zeta_3 \\
\end{array}
$$
see Figure \ref{pathfig}.
Measurement models of the reflective type were assumed, with 3 manifest ordinal reflective indicators for each latent variable
$$
X_{ih}=\,_X\lambda_{ih}\xi_i+\varepsilon_{ih},i=1,2,3,h=1,2,3 \quad \mathrm{and} \quad Y_{ih}=\,_Y\lambda_{ih}\eta_i+\delta_{ih},i=1,2,3,h=1,2,3.
$$
In order to take into account the presence of asymmetric distributions, latent variables $\xi_i, i=1,2,3$, were generated, in separate simulations, according both to the standard Normal distribution for all $\xi_i$ variables and Beta distributions with parameters $(\alpha=11,\beta=2)$ for $\xi_1$, $(\alpha=16,\beta=3)$ for $\xi_2$, $(\alpha=54,\beta=7)$ for $\xi_3$ which were then standardized. Theoretical asymmetry indices $\gamma_1=-0.9573$, $-0.7992$ and $-0.6043$ correspond to the three Beta distributions. Values of the asymmetry indices for the mobile phone data analyzed in Section \ref{TheMobilePhoneDataSet} are in the range $(-1.07,-0.22)$ except for one variable showing positive asymmetry. The model parameters were fixed to $\gamma_{11}=0.9,\gamma_{22}=0.5,\gamma_{23}=0.6,\beta_{21}=0.5$ and $\beta_{32}=0.6$. The $\lambda$ coefficients of measurement models were set to $0.8,0.9,0.95$. Both the variances of the error components $\zeta_i, i=1,2,3$ in the inner model and those pertaining errors in the measurement models were set to values ensuring the latent indicators $\eta_i, i=1,2,3$ and the manifest variables $X_{ih},i=1,2,h=1,2,3$ and $Y_{ih},i=1,2,3,h=1,2,3$ to have unit variance.

Manifest variables $X_{ih}$ and $Y_{ih}$ were rescaled according to the following rules
\begin{eqnarray*}
_{SCALED}X_{ih}&=&\frac{X_{ih}-\min(X_{ih})}{\max(X_{ih})-\min(X_{ih})+0.01}\cdot npoints+0.5 \\
_{SCALED}Y_{ih}&=&\frac{Y_{ih}-\min(Y_{ih})}{\max(Y_{ih})-\min(Y_{ih})+0.01}\cdot npoints+0.5
\end{eqnarray*}
with extrema computed over the sample realizations, being $npoints$ the desired number of points common to all items. Values were then rounded to obtain integer (ordinal) responses.

Simulations were performed by considering $4, 5, 7$ and 9 categories in the scales, which correspond to the situations commonly encountered in practice. We expect results from the PLS and the OPLS procedures to be quite similar in presence of 9 categories, since in this case polychoric correlations are close to their corresponding Pearson correlations.

500 replications for each instance, each with 250 observations were made.

To compare the performance of the two procedures we considered the empirical distribution of the inner model parameter estimate biases, see Tables \ref{biasdist4normal}-\ref{biasdist9}.

As expected \citep[see][]{Schneeweiss_1993} estimates obtained with the traditional PLS algorithm are negatively biased. Only for scales with $5, 7$ and 9 categories we can observe about 5\% trials with a small or negligible positive bias for Normal distributed latent variables. The bias gets more evident with decreasing number of scale points. The behaviour is common both to Normal and Beta situations. With the OPLS about 10\% simulations always present positive bias. Most percentage points of the bias distribution for the OPLS procedure are closer to 0 than with traditional PLS. Averages biases are again closer to 0 for the OPLS algorithm.

Percentage points and average values are very close for the two estimation procedures in case of a 9 point scale.

The ratio between the absolute biases observed in each trial with OPLS and the traditional PLS was considered, to better compare the two procedures. The distribution of the ratios is shown in the third sections of Tables \ref{biasdist4normal}-\ref{biasdist9} giving evidence that over 90\% trials have an absolute bias of the OPLS lower than the traditional PLS, when scales are characterized by 4 and 5 points. By comparing the 5\% and 95\% percentage points for the distributions of ratios of absolute biases in case of the Normal assumption with 4 point scales, we can observe the better behaviour of the OPLS: for parameter $\gamma_{22}$ we have 5\% and 95\% absolute ratios 0.0728 and 3.8032. According to the latter value 5\% trials have an absolute bias in OPLS estimates larger more than 3.8 times that of traditional PLS. According to the former value 5\% trials have an absolute bias of traditional PLS larger more than $1/0.0728=13.7$ times than OPLS.

Geometric means have been computed to summarize ratios between absolute biases of OPLS and traditional PLS and in all situations (except for $\gamma_{11}$, 9 points, Beta distribution) they are lower than 1. Their values increase with increasing number of scale points and get close to 1 in presence of scales with 9 points and asymmetric Beta distribution of the latent variables.

\begin{figure}
\begin{center}
\includegraphics[scale=.3]{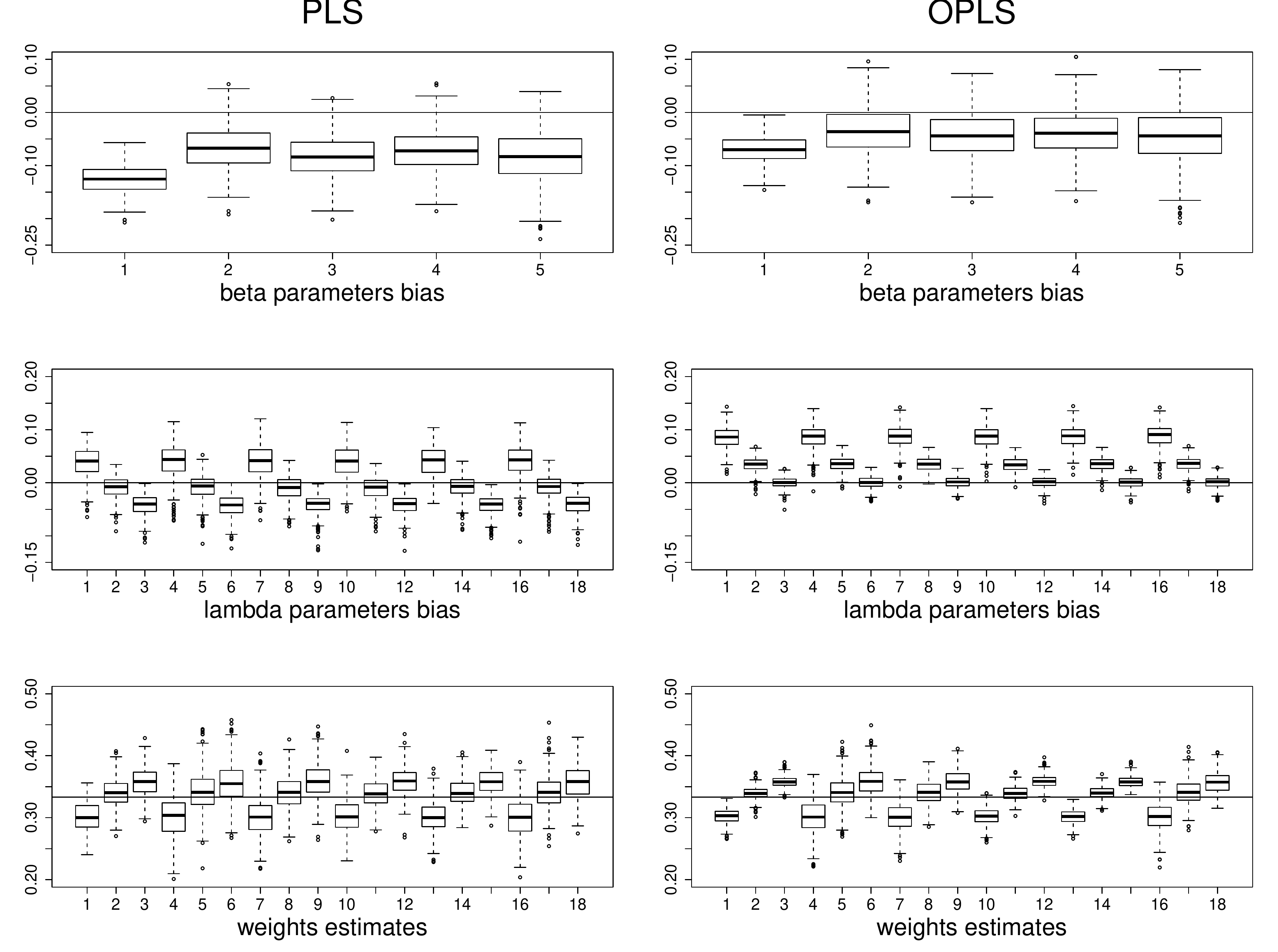}
\end{center}
\caption{Parameter estimates bias and weights distribution (4 points, Normal distribution)}
\label{normal4categoriesweights}
\end{figure}

\begin{figure}
\centering
\includegraphics[scale=.3]{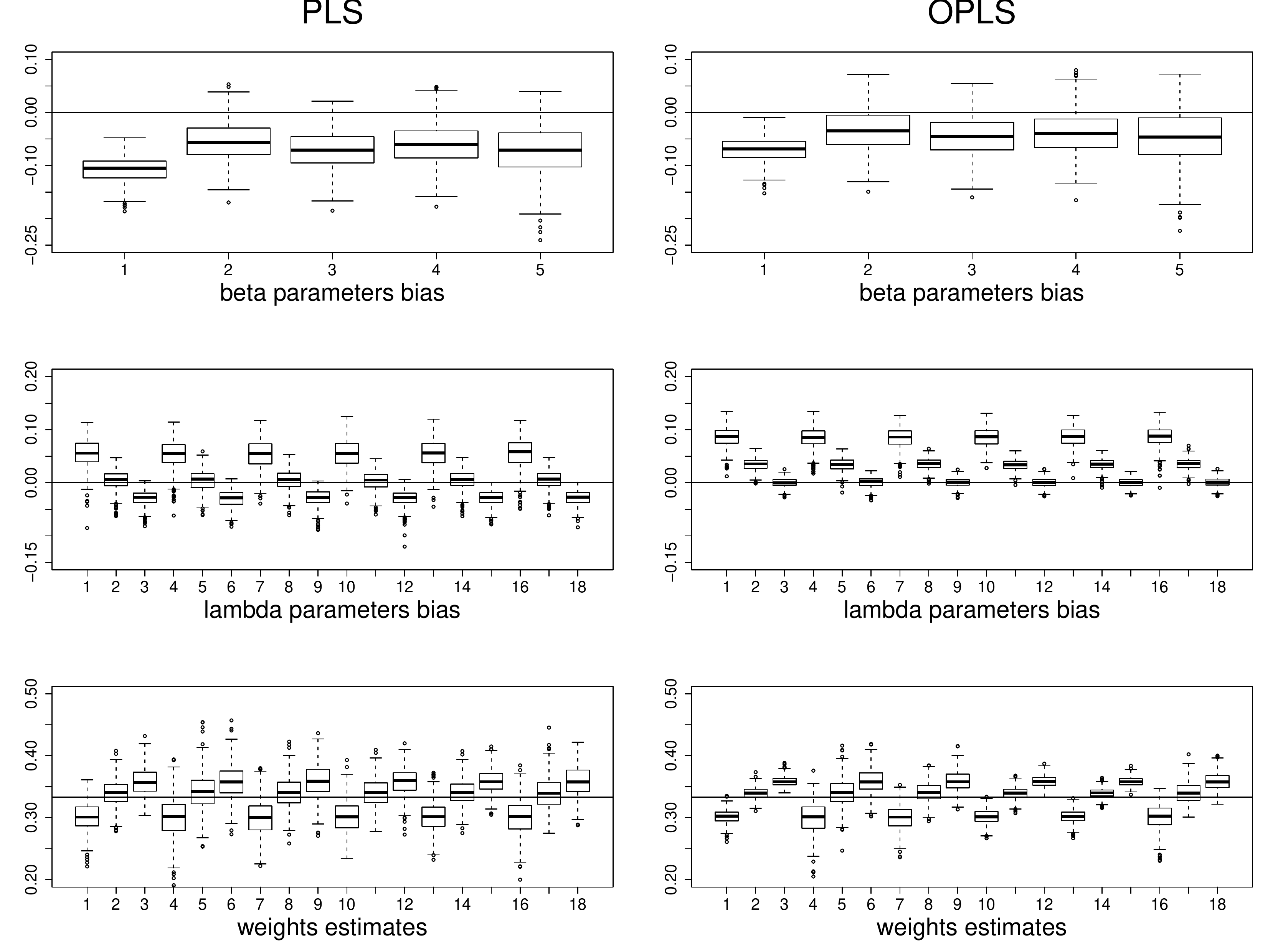}
\caption{\label{normal5categoriesweights}Parameter estimates bias and weights distribution (5 points, Normal distribution)}
\end{figure}

\begin{figure}
\centering
\includegraphics[scale=.3]{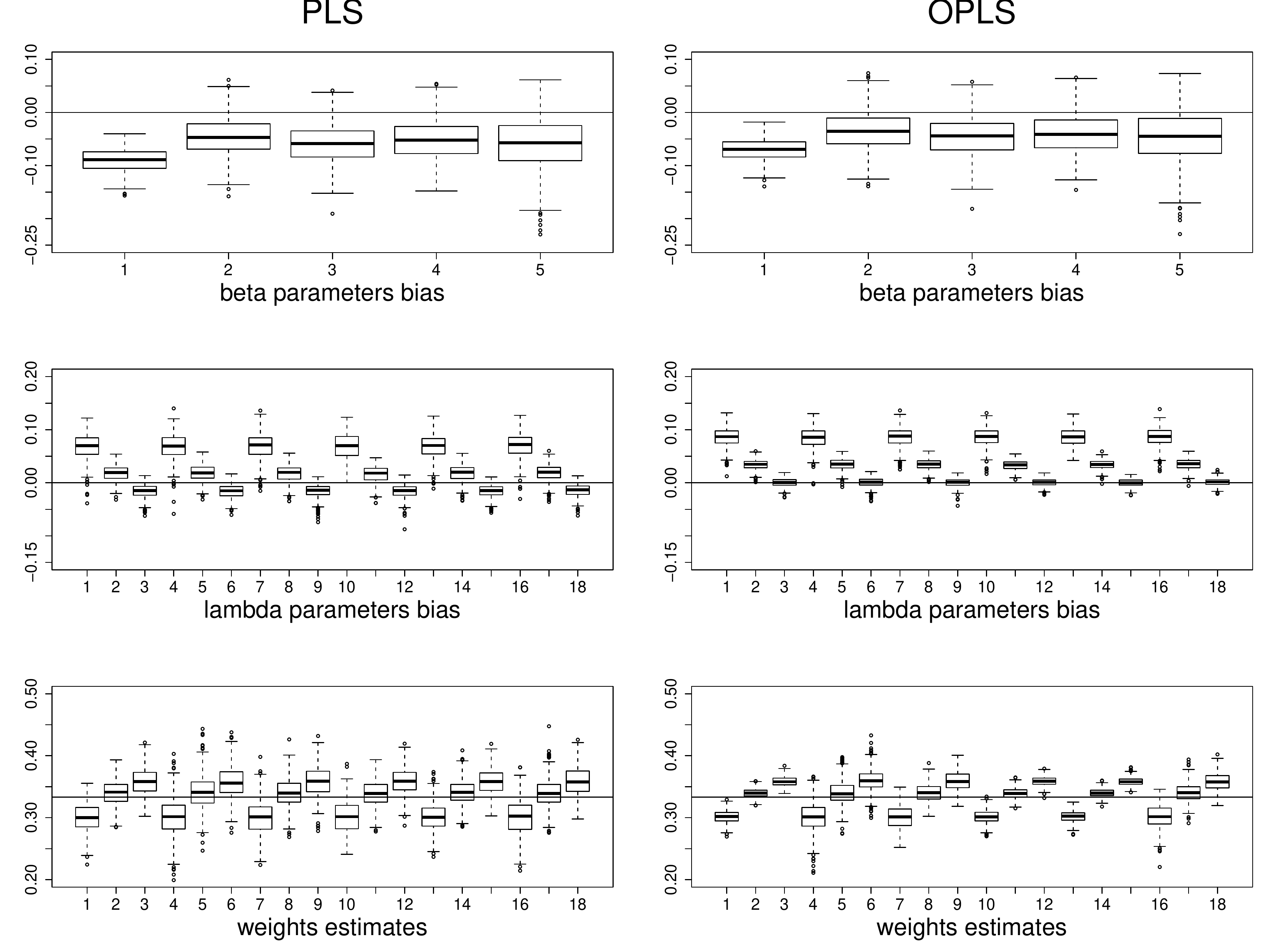}
\caption{\label{normal7categoriesweights}Parameter estimates bias and weights distribution (7 points, Normal distribution)}
\end{figure}

\begin{figure}
\begin{center}
\includegraphics[scale=.3]{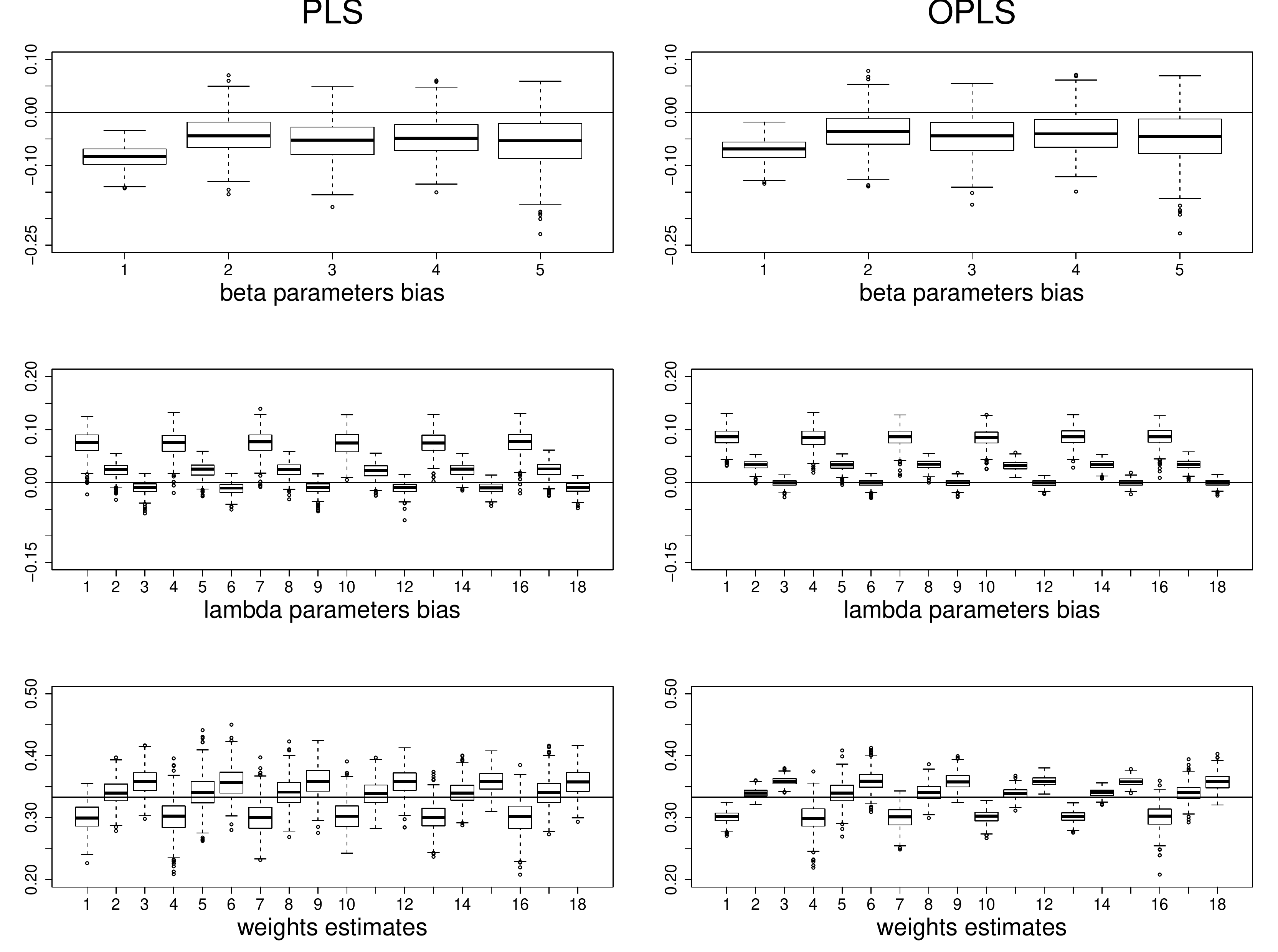}
\end{center}
\caption{Parameter estimates bias and weights distribution (9 points, Normal distribution)}
\label{normal9categoriesweights}
\end{figure}

\begin{figure}
\begin{center}
\includegraphics[scale=.3]{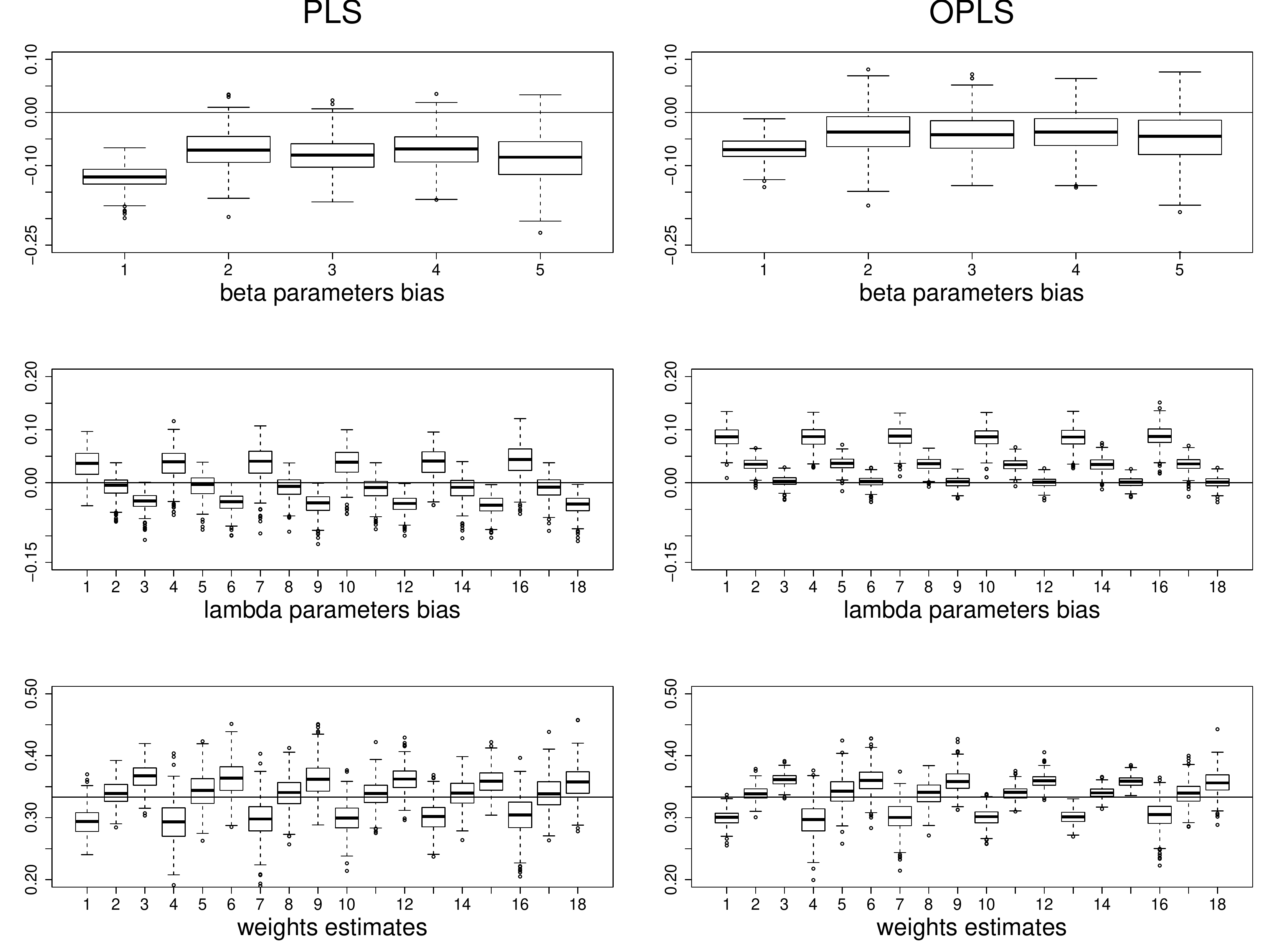}
\end{center}
\caption{Parameter estimates bias and weights distribution (4 categories, Beta distribution)}
\label{beta4categoriesweights}
\end{figure}

\begin{figure}
\centering
\includegraphics[scale=.3]{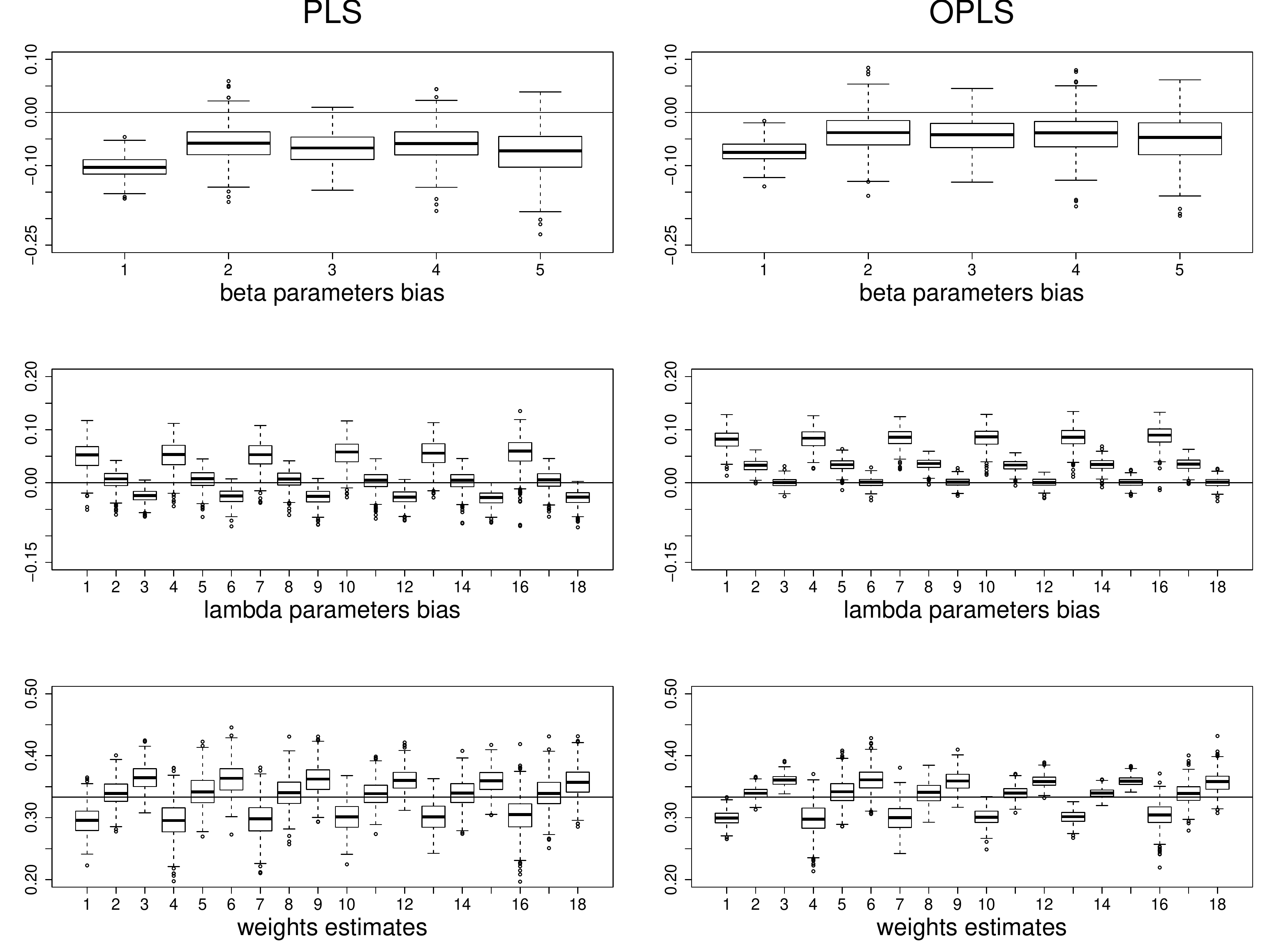}
\caption{\label{beta5categoriesweights}Parameter estimates bias and weights distribution (5 categories, Beta distribution)}
\end{figure}

\begin{figure}
\centering
\includegraphics[scale=.3]{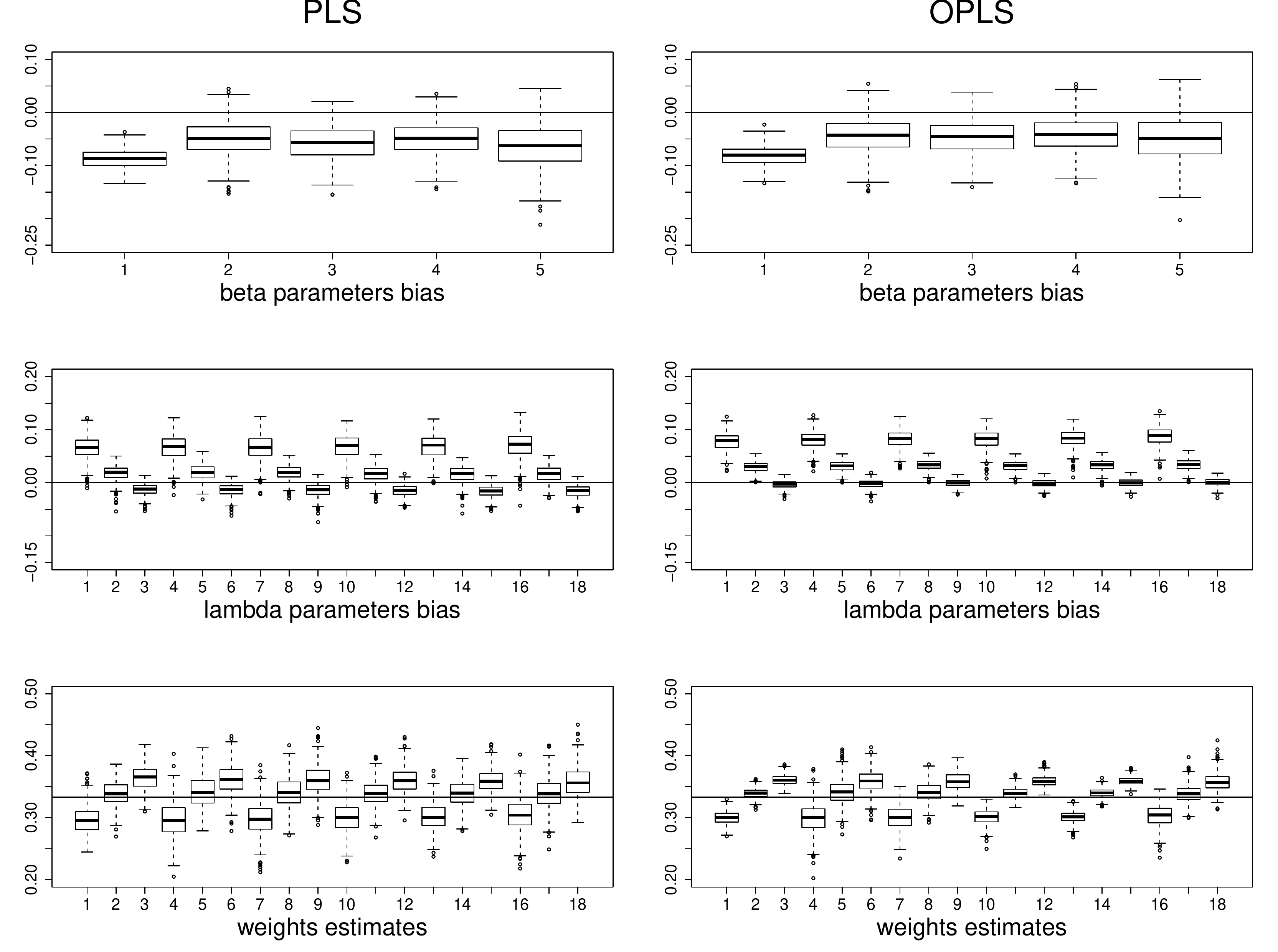}
\caption{\label{beta7categoriesweights}Parameter estimates bias and weights distribution (7 categories, Beta distribution)}
\end{figure}

\begin{figure}
\begin{center}
\includegraphics[scale=.3]{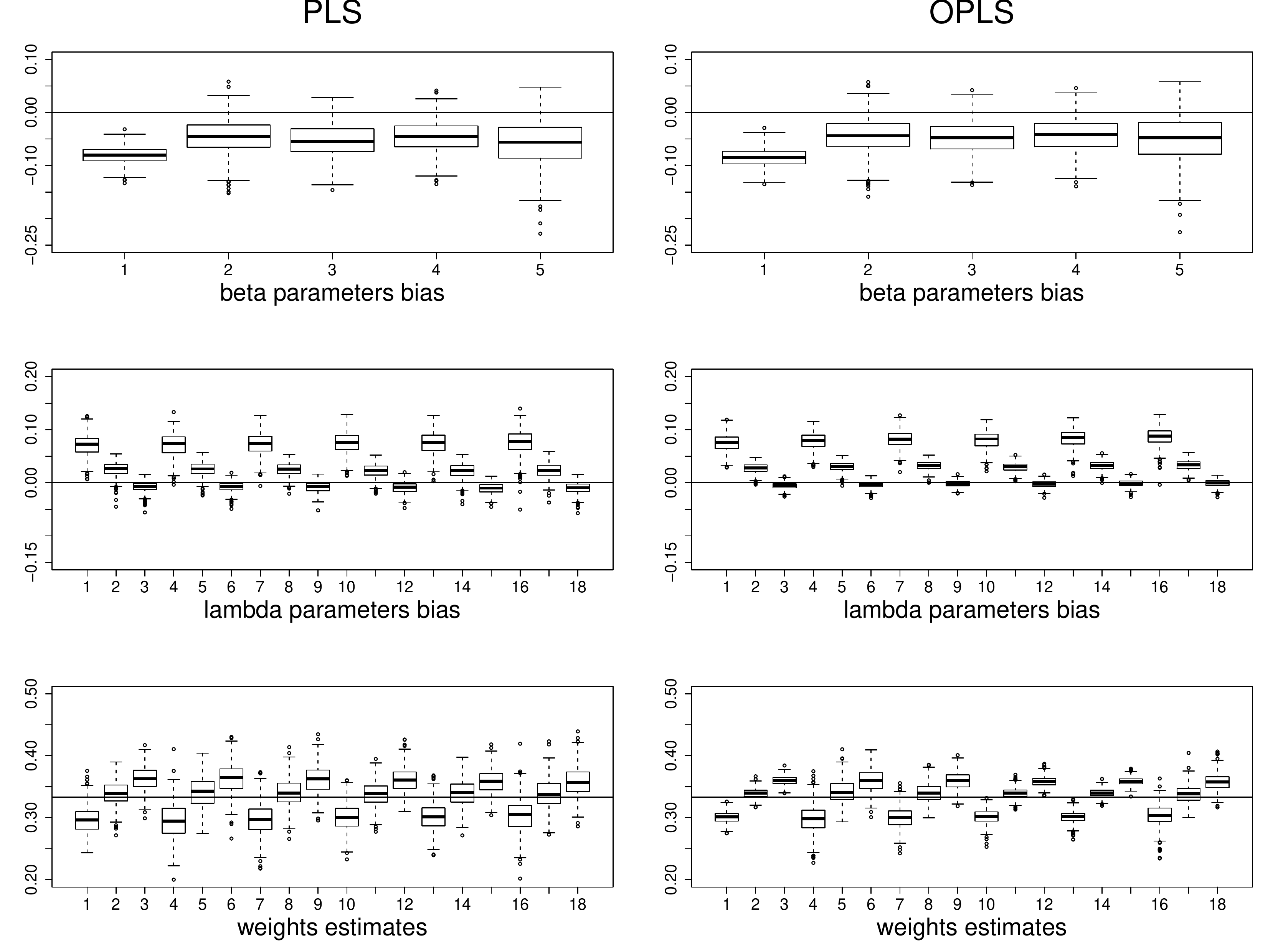}
\end{center}
\caption{Parameter estimates bias and weights distribution (9 categories, Beta distribution)}
\label{beta9categoriesweights}
\end{figure}

In Section \ref{EffectsofthebiasofthePLSalgorithmoncategoricalPLS} we reminded how the reduction in the bias attained by OPLS, pertaining the inner model parameter estimates, can have as a drawback an increase in the bias of the outer model $\lambda$ parameter estimates. The bias is evident if we examine Figures \ref{normal4categoriesweights}-\ref{beta9categoriesweights} which report Box \& Whiskers plots for the distribution of the bias of the coefficients estimates $\beta_{ij}$ and $\lambda$ from their theoretical values and the distribution of the weights $w_{ij}$ under the Normal and Beta assumptions for scales with 4 and 9 points.

However, as we have already remarked, the role played by outer parameters is less important than that of the inner model parameters: when making decisions based on PLS results the weights $w_{ij}$ are used instead of outer parameters; we remember that PLS define each latent variable as a 'composite' of its manifest indicators, see (\ref{boaricantaluppi.equation100}) and (\ref{equation120}), and the weights give information about the strength of the relationship of each 'composite' across its manifest indicators. According to the Box \& Whiskers Plots the estimates of the weights seem to be always characterized by a lower variability (interquartile range) when obtained with the OPLS algorithm.

\begin{table}
\caption{Bias distribution of the inner model parameter estimates (4 points, \label{biasdist4normal} Normal distribution) obtained with PLS and OPLS and distribution of the ratio between absolute values of the biases: percentage points, means and standard deviations}
\small
\begin{center}
\begin{tabular}{lrrrrrrrrr}
  \hline
 & 5\% & 10\% & 25\% & 50\% & 75\% & 90\% & 95\% & mean & sd \\ 
  \hline
& \multicolumn{9}{l}{PLS} \\
$\gamma_{11}=0.9$ & -0.166 & -0.158 & -0.144 & -0.125 & -0.107 & -0.094 & -0.087 & -0.126 & 0.025 \\ 
  $\gamma_{22}=0.5$ & -0.128 & -0.118 & -0.095 & -0.067 & -0.039 & -0.019 & -0.004 & -0.068 & 0.039 \\ 
  $\gamma_{23}=0.6$ & -0.147 & -0.131 & -0.110 & -0.084 & -0.056 & -0.033 & -0.021 & -0.083 & 0.038 \\ 
  $\beta_{21}=0.5$ & -0.131 & -0.119 & -0.098 & -0.072 & -0.046 & -0.023 & -0.010 & -0.072 & 0.038 \\ 
  $\beta_{32}=0.6$ & -0.164 & -0.149 & -0.115 & -0.083 & -0.050 & -0.022 & -0.006 & -0.084 & 0.049 \\ 
& \multicolumn{9}{l}{OPLS} \\
$\gamma_{11}=0.9$ & -0.111 & -0.103 & -0.087 & -0.070 & -0.052 & -0.039 & -0.027 & -0.070 & 0.025 \\ 
  $\gamma_{22}=0.5$ & -0.101 & -0.090 & -0.065 & -0.036 & -0.004 & 0.019 & 0.035 & -0.035 & 0.042 \\ 
  $\gamma_{23}=0.6$ & -0.111 & -0.095 & -0.072 & -0.044 & -0.014 & 0.009 & 0.023 & -0.044 & 0.042 \\ 
  $\beta_{21}=0.5$ & -0.103 & -0.091 & -0.067 & -0.039 & -0.011 & 0.016 & 0.031 & -0.039 & 0.042 \\ 
  $\beta_{32}=0.6$ & -0.138 & -0.111 & -0.077 & -0.044 & -0.010 & 0.020 & 0.036 & -0.046 & 0.052 \\ 
& \multicolumn{7}{l}{Ratio of absolute biases OPLS over PLS} & \multicolumn{2}{l}{geometric mean} \\
$\gamma_{11}=0.9$ & 0.329 & 0.392 & 0.465 & 0.557 & 0.613 & 0.666 & 0.693 & 0.522 &  \\ 
  $\gamma_{22}=0.5$ & 0.073 & 0.166 & 0.376 & 0.594 & 0.755 & 1.090 & 3.803 & 0.531 &  \\ 
  $\gamma_{23}=0.6$ & 0.113 & 0.182 & 0.385 & 0.577 & 0.697 & 0.792 & 0.982 & 0.483 &  \\ 
  $\beta_{21}=0.5$ & 0.100 & 0.207 & 0.414 & 0.621 & 0.747 & 0.914 & 2.559 & 0.543 &  \\ 
  $\beta_{32}=0.6$ & 0.112 & 0.244 & 0.436 & 0.606 & 0.736 & 0.911 & 3.437 & 0.575 &  \\ 
\hline
\end{tabular}
\end{center}
\end{table}

\begin{table}
\caption{Bias distribution of the inner model parameter estimates (5 points, \label{biasdist5normal} Normal distribution) obtained with PLS and OPLS and distribution of the ratio between absolute values of the biases: percentage points, means and standard deviations}
\small
\begin{center}
\begin{tabular}{lrrrrrrrrr}
  \hline
 & 5\% & 10\% & 25\% & 50\% & 75\% & 90\% & 95\% & mean & sd \\ 
  \hline
& \multicolumn{9}{l}{PLS} \\
$\gamma_{11}=0.9$ & -0.145 & -0.136 & -0.123 & -0.105 & -0.092 & -0.078 & -0.070 & -0.107 & 0.023 \\ 
  $\gamma_{22}=0.5$ & -0.120 & -0.106 & -0.079 & -0.057 & -0.029 & -0.008 & 0.004 & -0.056 & 0.037 \\ 
  $\gamma_{23}=0.6$ & -0.126 & -0.111 & -0.095 & -0.071 & -0.046 & -0.027 & -0.015 & -0.071 & 0.035 \\ 
  $\beta_{21}=0.5$ & -0.114 & -0.104 & -0.086 & -0.060 & -0.035 & -0.013 & 0.010 & -0.060 & 0.038 \\ 
  $\beta_{32}=0.6$ & -0.150 & -0.134 & -0.103 & -0.071 & -0.039 & -0.011 & 0.007 & -0.072 & 0.048 \\ 
& \multicolumn{9}{l}{OPLS} \\
$\gamma_{11}=0.9$ & -0.110 & -0.098 & -0.085 & -0.069 & -0.054 & -0.043 & -0.034 & -0.070 & 0.023 \\ 
  $\gamma_{22}=0.5$ & -0.100 & -0.088 & -0.060 & -0.035 & -0.005 & 0.018 & 0.027 & -0.034 & 0.039 \\ 
  $\gamma_{23}=0.6$ & -0.103 & -0.090 & -0.071 & -0.046 & -0.018 & 0.003 & 0.014 & -0.045 & 0.036 \\ 
  $\beta_{21}=0.5$ & -0.095 & -0.085 & -0.066 & -0.040 & -0.013 & 0.010 & 0.036 & -0.038 & 0.040 \\ 
  $\beta_{32}=0.6$ & -0.131 & -0.110 & -0.079 & -0.046 & -0.010 & 0.017 & 0.033 & -0.046 & 0.049 \\ 
& \multicolumn{7}{l}{Ratio of absolute biases OPLS over PLS} & \multicolumn{2}{l}{geometric mean} \\
$\gamma_{11}=0.9$ & 0.459 & 0.511 & 0.590 & 0.651 & 0.704 & 0.751 & 0.772 & 0.629 &  \\ 
  $\gamma_{22}=0.5$ & 0.107 & 0.220 & 0.509 & 0.700 & 0.823 & 1.670 & 4.287 & 0.641 &  \\ 
  $\gamma_{23}=0.6$ & 0.164 & 0.262 & 0.499 & 0.667 & 0.764 & 0.848 & 0.945 & 0.585 &  \\ 
  $\beta_{21}=0.5$ & 0.147 & 0.270 & 0.531 & 0.704 & 0.815 & 0.925 & 2.136 & 0.628 &  \\ 
  $\beta_{32}=0.6$ & 0.165 & 0.250 & 0.527 & 0.703 & 0.817 & 1.676 & 3.410 & 0.670 &  \\ 
\hline
\end{tabular}
\end{center}
\end{table}

\begin{table}
\caption{Bias distribution of the inner model parameter estimates (7 points, \label{biasdist7normal} Normal distribution) obtained with PLS and OPLS and distribution of the ratio between absolute values of the biases: percentage points, means and standard deviations}
\small
\begin{center}
\begin{tabular}{lrrrrrrrrr}
  \hline
 & 5\% & 10\% & 25\% & 50\% & 75\% & 90\% & 95\% & mean & sd \\ 
  \hline
& \multicolumn{9}{l}{PLS} \\
$\gamma_{11}=0.9$ & -0.128 & -0.117 & -0.105 & -0.089 & -0.074 & -0.063 & -0.055 & -0.090 & 0.022 \\ 
  $\gamma_{22}=0.5$ & -0.109 & -0.097 & -0.069 & -0.047 & -0.022 & -0.000 & 0.012 & -0.047 & 0.037 \\ 
  $\gamma_{23}=0.6$ & -0.115 & -0.103 & -0.084 & -0.059 & -0.035 & -0.013 & -0.002 & -0.059 & 0.035 \\ 
  $\beta_{21}=0.5$ & -0.107 & -0.098 & -0.077 & -0.052 & -0.027 & -0.004 & 0.014 & -0.051 & 0.036 \\ 
  $\beta_{32}=0.6$ & -0.140 & -0.118 & -0.091 & -0.057 & -0.025 & -0.001 & 0.012 & -0.059 & 0.048 \\ 
& \multicolumn{9}{l}{OPLS} \\
$\gamma_{11}=0.9$ & -0.107 & -0.099 & -0.084 & -0.070 & -0.055 & -0.044 & -0.037 & -0.070 & 0.021 \\ 
  $\gamma_{22}=0.5$ & -0.098 & -0.088 & -0.059 & -0.035 & -0.010 & 0.012 & 0.025 & -0.035 & 0.038 \\ 
  $\gamma_{23}=0.6$ & -0.102 & -0.091 & -0.071 & -0.044 & -0.020 & 0.002 & 0.012 & -0.045 & 0.037 \\ 
  $\beta_{21}=0.5$ & -0.095 & -0.089 & -0.066 & -0.041 & -0.014 & 0.007 & 0.027 & -0.040 & 0.037 \\ 
  $\beta_{32}=0.6$ & -0.131 & -0.104 & -0.077 & -0.045 & -0.011 & 0.012 & 0.024 & -0.045 & 0.048 \\ 
& \multicolumn{7}{l}{Ratio of absolute biases OPLS over PLS} & \multicolumn{2}{l}{geometric mean} \\
$\gamma_{11}=0.9$ & 0.617 & 0.667 & 0.725 & 0.779 & 0.824 & 0.867 & 0.889 & 0.764 &  \\ 
  $\gamma_{22}=0.5$ & 0.296 & 0.436 & 0.664 & 0.813 & 0.915 & 1.393 & 2.320 & 0.789 &  \\ 
  $\gamma_{23}=0.6$ & 0.277 & 0.443 & 0.657 & 0.794 & 0.877 & 0.959 & 1.635 & 0.733 &  \\ 
  $\beta_{21}=0.5$ & 0.283 & 0.420 & 0.683 & 0.845 & 0.904 & 1.242 & 1.785 & 0.767 &  \\ 
  $\beta_{32}=0.6$ & 0.272 & 0.462 & 0.679 & 0.827 & 0.910 & 1.761 & 3.156 & 0.840 &  \\ 
\hline
\end{tabular}
\end{center}
\end{table}

\begin{table}
\caption{Bias distribution of the inner model parameter estimates (9 points, \label{biasdist9normal} Normal distribution) obtained with PLS and OPLS and distribution of the ratio between absolute values of the biases: percentage points, means and standard deviations}
\small
\begin{center}
\begin{tabular}{lrrrrrrrrr}
  \hline
 & 5\% & 10\% & 25\% & 50\% & 75\% & 90\% & 95\% & mean & sd \\ 
  \hline
& \multicolumn{9}{l}{PLS} \\
$\gamma_{11}=0.9$ & -0.119 & -0.112 & -0.098 & -0.082 & -0.069 & -0.057 & -0.050 & -0.084 & 0.021 \\ 
  $\gamma_{22}=0.5$ & -0.104 & -0.092 & -0.066 & -0.044 & -0.018 & 0.001 & 0.015 & -0.043 & 0.036 \\ 
  $\gamma_{23}=0.6$ & -0.113 & -0.104 & -0.079 & -0.052 & -0.027 & -0.009 & 0.005 & -0.054 & 0.037 \\ 
  $\beta_{21}=0.5$ & -0.106 & -0.095 & -0.072 & -0.048 & -0.023 & -0.001 & 0.016 & -0.047 & 0.037 \\ 
  $\beta_{32}=0.6$ & -0.133 & -0.117 & -0.087 & -0.053 & -0.021 & 0.003 & 0.015 & -0.056 & 0.046 \\ 
& \multicolumn{9}{l}{OPLS} \\
$\gamma_{11}=0.9$ & -0.106 & -0.099 & -0.085 & -0.068 & -0.056 & -0.044 & -0.038 & -0.070 & 0.021 \\ 
  $\gamma_{22}=0.5$ & -0.099 & -0.085 & -0.060 & -0.036 & -0.011 & 0.009 & 0.024 & -0.036 & 0.037 \\ 
  $\gamma_{23}=0.6$ & -0.105 & -0.096 & -0.071 & -0.044 & -0.019 & 0.000 & 0.015 & -0.045 & 0.037 \\ 
  $\beta_{21}=0.5$ & -0.100 & -0.088 & -0.066 & -0.040 & -0.013 & 0.008 & 0.025 & -0.040 & 0.038 \\ 
  $\beta_{32}=0.6$ & -0.126 & -0.108 & -0.077 & -0.045 & -0.013 & 0.011 & 0.023 & -0.048 & 0.047 \\ 
& \multicolumn{7}{l}{Ratio of absolute biases OPLS over PLS} & \multicolumn{2}{l}{geometric mean} \\
$\gamma_{11}=0.9$ & 0.688 & 0.739 & 0.789 & 0.844 & 0.894 & 0.936 & 0.956 & 0.832 &  \\ 
  $\gamma_{22}=0.5$ & 0.316 & 0.508 & 0.754 & 0.884 & 0.971 & 1.205 & 1.672 & 0.837 &  \\ 
  $\gamma_{23}=0.6$ & 0.293 & 0.539 & 0.731 & 0.867 & 0.939 & 1.025 & 1.439 & 0.809 &  \\ 
  $\beta_{21}=0.5$ & 0.305 & 0.491 & 0.753 & 0.889 & 0.955 & 1.149 & 1.560 & 0.833 &  \\ 
  $\beta_{32}=0.6$ & 0.370 & 0.535 & 0.784 & 0.884 & 0.957 & 1.410 & 2.006 & 0.867 &  \\ 
\hline
\end{tabular}
\end{center}
\end{table}

\begin{table}
\caption{Bias distribution of the inner model parameter estimates (4 points, \label{biasdist4} Beta distribution) obtained with PLS and OPLS and distribution of the ratio between absolute values of the biases: percentage points, means and standard deviations}
\small
\begin{center}
\begin{tabular}{lrrrrrrrrr}
  \hline
 & 5\% & 10\% & 25\% & 50\% & 75\% & 90\% & 95\% & mean & sd \\ 
  \hline
& \multicolumn{9}{l}{PLS} \\
$\gamma_{11}=0.9$ & -0.160 & -0.151 & -0.135 & -0.122 & -0.107 & -0.094 & -0.088 & -0.122 & 0.022 \\ 
  $\gamma_{22}=0.5$ & -0.130 & -0.116 & -0.094 & -0.071 & -0.045 & -0.027 & -0.012 & -0.070 & 0.036 \\ 
  $\gamma_{23}=0.6$ & -0.136 & -0.125 & -0.103 & -0.080 & -0.059 & -0.038 & -0.025 & -0.081 & 0.034 \\ 
  $\beta_{21}=0.5$ & -0.128 & -0.116 & -0.093 & -0.069 & -0.046 & -0.025 & -0.015 & -0.070 & 0.035 \\ 
  $\beta_{32}=0.6$ & -0.152 & -0.139 & -0.117 & -0.084 & -0.055 & -0.030 & -0.013 & -0.086 & 0.044 \\ 
& \multicolumn{9}{l}{OPLS} \\
$\gamma_{11}=0.9$ & -0.106 & -0.097 & -0.083 & -0.070 & -0.054 & -0.043 & -0.035 & -0.070 & 0.021 \\ 
  $\gamma_{22}=0.5$ & -0.102 & -0.089 & -0.064 & -0.037 & -0.008 & 0.011 & 0.023 & -0.038 & 0.040 \\ 
  $\gamma_{23}=0.6$ & -0.103 & -0.091 & -0.067 & -0.042 & -0.016 & 0.006 & 0.023 & -0.042 & 0.038 \\ 
  $\beta_{21}=0.5$ & -0.099 & -0.088 & -0.062 & -0.037 & -0.012 & 0.015 & 0.026 & -0.038 & 0.038 \\ 
  $\beta_{32}=0.6$ & -0.118 & -0.103 & -0.079 & -0.045 & -0.014 & 0.011 & 0.028 & -0.047 & 0.046 \\ 
& \multicolumn{7}{l}{Ratio of absolute biases OPLS over PLS} & \multicolumn{2}{l}{geometric mean} \\
$\gamma_{11}=0.9$ & 0.371 & 0.426 & 0.501 & 0.575 & 0.629 & 0.681 & 0.704 & 0.549 &  \\ 
  $\gamma_{22}=0.5$ & 0.075 & 0.142 & 0.357 & 0.598 & 0.738 & 0.850 & 2.108 & 0.488 &  \\ 
  $\gamma_{23}=0.6$ & 0.101 & 0.189 & 0.384 & 0.559 & 0.692 & 0.781 & 0.888 & 0.454 &  \\ 
  $\beta_{21}=0.5$ & 0.127 & 0.213 & 0.415 & 0.598 & 0.741 & 0.842 & 1.621 & 0.522 &  \\ 
  $\beta_{32}=0.6$ & 0.099 & 0.226 & 0.397 & 0.590 & 0.720 & 0.828 & 2.197 & 0.503 &  \\ 
\hline
\end{tabular}
\end{center}
\end{table}

\begin{table}
\caption{Bias distribution of the inner model parameter estimates (5 points, \label{biasdist5} Beta distribution) obtained with PLS and OPLS and distribution of the ratio between absolute values of the biases: percentage points, means and standard deviations}
\small
\begin{center}
\begin{tabular}{lrrrrrrrrr}
  \hline
 & 5\% & 10\% & 25\% & 50\% & 75\% & 90\% & 95\% & mean & sd \\ 
  \hline
& \multicolumn{9}{l}{PLS} \\
$\gamma_{11}=0.9$ & -0.135 & -0.126 & -0.116 & -0.104 & -0.089 & -0.079 & -0.072 & -0.103 & 0.019 \\ 
  $\gamma_{22}=0.5$ & -0.114 & -0.099 & -0.079 & -0.058 & -0.037 & -0.014 & 0.001 & -0.057 & 0.034 \\ 
  $\gamma_{23}=0.6$ & -0.118 & -0.107 & -0.088 & -0.067 & -0.047 & -0.028 & -0.017 & -0.067 & 0.031 \\ 
  $\beta_{21}=0.5$ & -0.115 & -0.102 & -0.080 & -0.059 & -0.037 & -0.016 & 0.001 & -0.059 & 0.034 \\ 
  $\beta_{32}=0.6$ & -0.145 & -0.131 & -0.103 & -0.072 & -0.045 & -0.013 & 0.000 & -0.073 & 0.044 \\ 
& \multicolumn{9}{l}{OPLS} \\
$\gamma_{11}=0.9$ & -0.106 & -0.097 & -0.087 & -0.075 & -0.060 & -0.047 & -0.041 & -0.074 & 0.020 \\ 
  $\gamma_{22}=0.5$ & -0.099 & -0.084 & -0.061 & -0.038 & -0.015 & 0.010 & 0.025 & -0.038 & 0.037 \\ 
  $\gamma_{23}=0.6$ & -0.100 & -0.087 & -0.066 & -0.042 & -0.021 & -0.001 & 0.010 & -0.043 & 0.033 \\ 
  $\beta_{21}=0.5$ & -0.099 & -0.086 & -0.065 & -0.038 & -0.017 & 0.007 & 0.022 & -0.039 & 0.037 \\ 
  $\beta_{32}=0.6$ & -0.123 & -0.108 & -0.079 & -0.047 & -0.020 & 0.014 & 0.027 & -0.048 & 0.046 \\ 
& \multicolumn{7}{l}{Ratio of absolute biases OPLS over PLS} & \multicolumn{2}{l}{geometric mean} \\
$\gamma_{11}=0.9$ & 0.537 & 0.581 & 0.653 & 0.722 & 0.780 & 0.827 & 0.857 & 0.704 &  \\ 
  $\gamma_{22}=0.5$ & 0.199 & 0.325 & 0.558 & 0.724 & 0.851 & 0.987 & 2.509 & 0.686 &  \\ 
  $\gamma_{23}=0.6$ & 0.146 & 0.259 & 0.502 & 0.664 & 0.789 & 0.856 & 0.926 & 0.564 &  \\ 
  $\beta_{21}=0.5$ & 0.236 & 0.353 & 0.546 & 0.719 & 0.832 & 0.962 & 2.579 & 0.687 &  \\ 
  $\beta_{32}=0.6$ & 0.221 & 0.367 & 0.560 & 0.720 & 0.824 & 1.124 & 3.351 & 0.706 &  \\ 
\hline
\end{tabular}
\end{center}
\end{table}

\begin{table}
\caption{Bias distribution of the inner model parameter estimates (7 points, \label{biasdist7} Beta distribution) obtained with PLS and OPLS and distribution of the ratio between absolute values of the biases: percentage points, means and standard deviations}
\small
\begin{center}
\begin{tabular}{lrrrrrrrrr}
  \hline
 & 5\% & 10\% & 25\% & 50\% & 75\% & 90\% & 95\% & mean & sd \\ 
  \hline
& \multicolumn{9}{l}{PLS} \\
$\gamma_{11}=0.9$ & -0.117 & -0.111 & -0.099 & -0.087 & -0.075 & -0.066 & -0.061 & -0.088 & 0.017 \\ 
  $\gamma_{22}=0.5$ & -0.101 & -0.086 & -0.069 & -0.049 & -0.027 & -0.006 & 0.003 & -0.049 & 0.033 \\ 
  $\gamma_{23}=0.6$ & -0.107 & -0.095 & -0.080 & -0.056 & -0.035 & -0.019 & -0.009 & -0.057 & 0.031 \\ 
  $\beta_{21}=0.5$ & -0.104 & -0.094 & -0.070 & -0.048 & -0.029 & -0.008 & 0.000 & -0.050 & 0.032 \\ 
  $\beta_{32}=0.6$ & -0.135 & -0.122 & -0.091 & -0.063 & -0.035 & -0.005 & 0.010 & -0.063 & 0.044 \\ 
& \multicolumn{9}{l}{OPLS} \\
$\gamma_{11}=0.9$ & -0.114 & -0.107 & -0.094 & -0.080 & -0.069 & -0.059 & -0.053 & -0.082 & 0.019 \\ 
  $\gamma_{22}=0.5$ & -0.099 & -0.082 & -0.065 & -0.043 & -0.021 & 0.003 & 0.013 & -0.042 & 0.034 \\ 
  $\gamma_{23}=0.6$ & -0.100 & -0.087 & -0.069 & -0.045 & -0.024 & -0.005 & 0.004 & -0.047 & 0.032 \\ 
  $\beta_{21}=0.5$ & -0.098 & -0.087 & -0.063 & -0.041 & -0.020 & 0.000 & 0.010 & -0.042 & 0.033 \\ 
  $\beta_{32}=0.6$ & -0.122 & -0.109 & -0.078 & -0.049 & -0.019 & 0.009 & 0.021 & -0.050 & 0.044 \\ 
& \multicolumn{7}{l}{Ratio of absolute biases OPLS over PLS} & \multicolumn{2}{l}{geometric mean} \\
$\gamma_{11}=0.9$ & 0.781 & 0.816 & 0.868 & 0.932 & 0.997 & 1.048 & 1.098 & 0.926 &  \\ 
  $\gamma_{22}=0.5$ & 0.415 & 0.585 & 0.792 & 0.908 & 1.018 & 1.186 & 1.778 & 0.876 &  \\ 
  $\gamma_{23}=0.6$ & 0.293 & 0.463 & 0.700 & 0.831 & 0.919 & 1.013 & 1.154 & 0.751 &  \\ 
  $\beta_{21}=0.5$ & 0.351 & 0.456 & 0.729 & 0.872 & 0.971 & 1.125 & 2.057 & 0.835 &  \\ 
  $\beta_{32}=0.6$ & 0.353 & 0.502 & 0.701 & 0.826 & 0.901 & 1.078 & 2.075 & 0.817 &  \\ 
\hline
\end{tabular}
\end{center}
\end{table}

\begin{table}
\caption{Bias distribution of the inner model parameter estimates (9 points, \label{biasdist9} Beta distribution) obtained with PLS and OPLS and distribution of the ratio between absolute values of the biases: percentage points, means and standard deviations}
\small
\begin{center}
\begin{tabular}{lrrrrrrrrr}
  \hline
 & 5\% & 10\% & 25\% & 50\% & 75\% & 90\% & 95\% & mean & sd \\ 
  \hline
& \multicolumn{9}{l}{PLS} \\
$\gamma_{11}=0.9$ & -0.107 & -0.101 & -0.091 & -0.080 & -0.070 & -0.059 & -0.053 & -0.080 & 0.016 \\ 
  $\gamma_{22}=0.5$ & -0.098 & -0.084 & -0.065 & -0.045 & -0.024 & -0.002 & 0.008 & -0.045 & 0.033 \\ 
  $\gamma_{23}=0.6$ & -0.102 & -0.092 & -0.073 & -0.054 & -0.031 & -0.011 & -0.004 & -0.053 & 0.030 \\ 
  $\beta_{21}=0.5$ & -0.097 & -0.087 & -0.065 & -0.045 & -0.025 & -0.004 & 0.008 & -0.045 & 0.032 \\ 
  $\beta_{32}=0.6$ & -0.127 & -0.116 & -0.086 & -0.056 & -0.028 & -0.004 & 0.014 & -0.058 & 0.043 \\ 
& \multicolumn{9}{l}{OPLS} \\
$\gamma_{11}=0.9$ & -0.115 & -0.110 & -0.097 & -0.085 & -0.073 & -0.061 & -0.056 & -0.085 & 0.018 \\ 
  $\gamma_{22}=0.5$ & -0.095 & -0.086 & -0.064 & -0.044 & -0.021 & 0.001 & 0.014 & -0.043 & 0.034 \\ 
  $\gamma_{23}=0.6$ & -0.098 & -0.087 & -0.069 & -0.048 & -0.027 & -0.006 & 0.004 & -0.048 & 0.031 \\ 
  $\beta_{21}=0.5$ & -0.097 & -0.085 & -0.064 & -0.042 & -0.021 & -0.001 & 0.010 & -0.042 & 0.032 \\ 
  $\beta_{32}=0.6$ & -0.121 & -0.108 & -0.079 & -0.048 & -0.019 & 0.004 & 0.019 & -0.050 & 0.043 \\ 
& \multicolumn{7}{l}{Ratio of absolute biases OPLS over PLS} & \multicolumn{2}{l}{geometric mean} \\
$\gamma_{11}=0.9$ & 0.885 & 0.926 & 0.984 & 1.061 & 1.132 & 1.205 & 1.244 & 1.056 &  \\ 
  $\gamma_{22}=0.5$ & 0.561 & 0.699 & 0.856 & 0.986 & 1.097 & 1.332 & 1.709 & 0.973 &  \\ 
  $\gamma_{23}=0.6$ & 0.470 & 0.658 & 0.823 & 0.922 & 1.028 & 1.202 & 1.763 & 0.914 &  \\ 
  $\beta_{21}=0.5$ & 0.467 & 0.661 & 0.829 & 0.945 & 1.070 & 1.286 & 1.747 & 0.918 &  \\ 
  $\beta_{32}=0.6$ & 0.381 & 0.569 & 0.778 & 0.884 & 0.949 & 1.084 & 1.463 & 0.833 &  \\ 
\hline
\end{tabular}
\end{center}
\end{table}


\section{Conclusions and Final Remarks}

A PLS algorithm dealing with variables on ordinal scales has been presented. The algorithm, OPLS, is based on the use of the polychoric correlation matrix and seems to perform better than the traditional PLS algorithm in presence of ordinal scales with a small number of point alternatives, by reducing the bias of the inner model parameter estimates.

A basic feature of PLS is the so-called soft modelling, requiring no distributional assumptions on the variables appearing in the structural equation model. With the OPLS algorithm the continuous variables underlying the categorical manifest indicators are considered multinormally distributed.
This can appear a strong assumption but, as \cite{Bartolomew_1996} observes, every distribution can be obtained as a transformation of the Normal one, which can suit most situations: for instance, in presence of a manifest variable with a negative asymmetric distribution, points on the right side of a scale will have the highest frequency and the underlying latent variable should also be of an asymmetric type, but transformation (\ref{ZBBCequation003}) will work anyway assigning larger intervals to the classes defined by the thresholds to which the highest points in the scale correspond.

Furthermore polychoric correlations are expected to overestimate real correlations when scales present some kind of asymmetry, but this can be regarded as a positive feature for the OPLS algorithm. This may represent a correction of the negative bias characterizing PLS algorithms with regard to the estimates of the inner model parameters (which are in some way linked to correlations across manifest variables).

The gain in the bias reduction is less evident for scales with an higher number of categories, for which polychoric correlation values are closer to Pearson's correlations. In these cases ordinal scales can be considered as they were of the interval type, possibly according to the so-called pragmatic approach to measurement \citep{Hand_2009}.

Increasing the number of the points of the scale can help the performance of the traditional PLS algorithm when the scale is interpreted as continuous, but it often happens that in presence of asymmetric distributions many points of the scale are characterized by a very low response frequency, since the number of points that respondents do effectively choose may be quite restricted. Thus the administered scale corresponds to a scale with a lower number of points and OPLS can anyway be useful in these situations.

Another important feature of the PLS predictive approach is the direct estimation of latent scores. With the OPLS algorithm we can estimate some thresholds for the latent variables, from which a 'category' indication for the ordinal latent variable follows according to one of the 3 estimation methods presented in Section \ref{Prediction}.

Simulations have been carried out to assess the properties of the algorithm also in presence of asymmetric distributions for latent variables and the bias characterizing the inner model parameter estimates obtained with the traditional PLS algorithm was reduced.

Further research will consider a comparison with the Optimal Scaling techniques \citep{MairLeeuw:2010} that were proposed within the PLS framework by \cite{Nappo} and \cite{LauroNappoGrassiaMiele}.

\bibliographystyle{model5-names}
\bibliography{Cantaluppi}

\newpage
~
\thispagestyle{empty}

\vspace{17.7cm}
\begin{center}
Finito di stampare nel mese di Dicembre 2012
\\presso la Redazione e composizione stampati 
\\Universit\`a Cattolica del Sacro Cuore

\vspace{9pt}
La Redazione ottempera agli obblighi previsti 
\\dalla L. 106/2004 e dal DPR 252/2006
\end{center}

\end{document}